\documentclass[%
superscriptaddress,
preprint, onecolumn,
preprintnumbers,
amsmath,amssymb,
aps,
floatfix,
]{revtex4-1}

\usepackage{graphicx}% Include figure files
\usepackage{bm}% bold math
\usepackage{hyperref}% add hypertext capabilities
\usepackage{amsmath}
\usepackage{amssymb}
\usepackage{makecell}
\usepackage{rotating}
\usepackage{tabularray}

\newcommand{\barphi}{\overline{\langle{\phi_f}\rangle}}

\begin{document}

\title{Solute Transport due to Periodic Loading in a Soft Porous Material}

\author{Matilde Fiori}
\affiliation{
    Department of Engineering Science, University of Oxford, Oxford, OX1 3PJ, UK }
    \affiliation{Institut de Mécanique Des Fluides de Toulouse, IMFT,
Université de Toulouse, CNRS, Toulouse, 31400, France }

\author{Satyajit Pramanik}
\affiliation{
    Department of Mathematics, Indian Institute of Technology Guwahati, Guwahati -- 781039, Assam, India }

\author{Christopher W. MacMinn}
\email{christopher.macminn@eng.ox.ac.uk}
\affiliation{
    Department of Engineering Science, University of Oxford, Oxford, OX1 3PJ, UK }

\date{\today}

\begin{abstract}
In soft porous media, deformation drives solute transport via the
intrinsic coupling between flow of the fluid and rearrangement of the
pore structure. Solute transport driven by periodic loading, in particular, can be of great relevance in applications including the geomechanics of contaminants in the subsurface and the biomechanics of nutrient transport in living tissues and scaffolds for tissue engineering. However, the basic features of this process have not previously been systematically investigated. Here, we fill this hole in the context of a 1D model problem. We do so by expanding the results from a companion study, in which we explored the poromechanics of periodic deformations, by introducing and analysing the impact of the resulting fluid and solid motion on solute transport. We first characterise the independent roles of the three main mechanisms of solute transport in porous media --- advection, molecular diffusion, and hydrodynamic dispersion --- by examining their impacts on the solute concentration profile during one loading cycle. We next explore the impact of the transport parameters, showing how these alter the relative importance of diffusion and dispersion. We then explore the loading parameters by considering a range of loading periods --- from slow to fast, relative to the poroelastic timescale --- and amplitudes --- from infinitesimal to large. We show that solute spreading over several loading cycle increases monotonically with amplitude, but is maximised for intermediate periods because of the increasing poromechanical localisation of the flow and deformation near the permeable boundary as the period decreases.
\end{abstract}

\maketitle

%This is how we can use Matilde's own \LaTeX command: 
%\begin{verbatim}
%    \matilde{x}
%\end{verbatim}
%produces $\matilde{x}$. 

\section{\label{Intro}Introduction}

Solutes spread and mix in deformable porous media in a variety of geomechanical, industrial, and biological contexts. In general, the transport of solutes in porous media is driven by molecular diffusion and by internal fluid flow. In soft porous media, the latter is strongly coupled to external mechanical loads  through rearrangement of the pore structure \citep[\textit{e.g.},][]{Mow1980, Lai1991,PREZIOSI1996, Li2004,  Franceschini2006, Moeendarbary2013, Ehlers2009, Vuong2015,  Borja2016}. In many cases, these loads are periodic; for example, compression due to surface loading can induce the spreading of contaminants in soils, exacerbating environmental harm and hindering remediation, while physiological loads can drive nutrient transport and waste removal in biological tissues, thus potentially playing an important role in cell growth and survival. In a companion study \citep{Fiori2023}, we examined the poromechanics of periodic loading over a wide range of loading periods and amplitudes. Here, we examine the implications of those results for solute transport.

At the continuum (Darcy) scale, which is the framework of this study, solute transport occurs through three fundamental mechanisms: advection, diffusion, and hydrodynamic dispersion \cite{bear72, Whitaker67, Saffman1959, Scheidegger1961, brenner1993, whitaker1999method, gelhar1993stochastic, DENTZ20111}. Diffusion in a porous medium is weaker than in a bulk fluid because of the tortuosity of the pore space \cite{bear72, ghanbarian2013, Tartakovsky}. Both advection and dispersion are driven by fluid flow. Advection is driven by the mean interstitial fluid velocity. Dispersion results from the pore-scale deviations from this Darcy-scale mean. In particular, dispersion is driven by two phenomena: (i) analogous to classical Taylor dispersion in a tube \cite{taylor1953dispersion, Brenner80, Marbach2019}, pore-scale velocity gradients smear solute profiles along the flow direction, inducing longitudinal spreading, and (ii) the morphology of the pore structure introduces chaotic variability in the fluid streamlines \cite{Datta2013, deAnna2013,Lester2013, lester_dentz_leborgne_2016, LESTER2016, Gouze21, Dentz2023, Kree2017, Souzy2020}, thus inducing both longitudinal and transverse spreading \cite{Scheidegger1961, GelharAxness, Gelhar92, DELGADO2007}. In soft porous media, therefore, deformation can enhance the transport of solutes directly by driving fluid flow, thus leading to advection and dispersion, and indirectly by distorting the pore space, and thus modifying both dispersion and diffusion.

Solute transport in deformable porous media has been studied in several different contexts. The impact of monotonic soil consolidation on transport has been studied extensively for its relevance to the management of landfills and other contaminated sediments, such as dredging or mining waste~\citep[\textit{e.g.},][]{Smith2000, Peters2002, alshawabkeh2006, Fox2007, fox2007b,arega-applmathmodelling-2008, Lewis12, zhang-awr-2012, Zhang2013, xie-envscipollutionres-2016, Pu2018, Bonazzi2021}. In that context, it is well known that consolidation enhances solute transport. Deformation has also been shown to increase mixing and reduce breakthrough time in the context of miscible viscous fingering~\cite{tran2020coupling}. The key feature introduced by periodic loading is the continuously fluctuating fluid flow, which can irreversibly modify diffusion and dispersion even when the macroscopic advective component is perfectly reversible. The role of periodic flow in enhancing solute transport and mixing has been studied in rigid and compressible 1D pore networks~\cite{Goldsztein2004, Claria2012}. In a poroelastic material, solute transport due to small periodic deformations has been explored across a range of parameters, including compressibility and forcing frequency, for semi-infinite homogeneous systems~\cite{Pool2016}, finite homogeneous systems~\cite{Bonazzi2021}, and finite heterogeneous systems~\cite{Trefry2019, Wu2020}. The latter two studies focus in particular on the combined role of poroelasticity, heterogeneity, and transient forcing in generating chaotic advection.

Periodic loading is also known to enhance the transport of nutrients in biological tissues~\citep[][]{Ferguson2004, gardiner2007, Witt2014, Schmidt2010, Zhang2011, DiDomenico2017,  LeZHANG2008430}. Similarly, periodic deformations are used to enhance the infiltration of solutes into hydrogels~\cite{Albro2008, Vaughan2013} and other scaffolds for tissue engineering~\cite{Mauck2003, Cortez2016, Kumar2018, Fan2016}, where the correlation between loading parameters, nutrient transport, and cell survival is of particular interest. Increasing the loading amplitude and/or decreasing the loading period induces a transition from diffusion-dominated to advection-dominated regimes~\cite{Urciuolo2008} and amplifies the role of hydrodynamic dispersion~\cite{Sengers2004}. Decreasing the loading period also leads to localisation of flow and deformation near permeable boundaries, resulting in larger velocities near the surface that promote external solute infiltration~\cite{gardiner2007, Urciuolo2008, DiDomenico2017, Vaughan2013}.

In general, despite the established role of hydrodynamic dispersion in driving the transport of solutes in porous media, dispersion is rarely included in biomechanical models (with the notable exception of Ref. \cite{Sengers2004}). One context where dispersion is widely agreed to be important is in brain microcirculation \cite{Kelley23}. In the vascular network within the brain, dispersion results from the shear-induced radial concentration gradients in single vessels~\citep[\textit{e.g.},][]{Marbach2019, Sharp19, berg2020, Troyesky21, bojarskaite-natcomms-2023} and the progressive bifurcation of vessels into smaller branches that can be modelled at the continuum scale as a porous material~\citep[\textit{e.g.},][]{zimmerman2020, Goirand21}.

Dispersion is typically neglected in the context of tissues and gels for two main reasons. First, fluid flow is often assumed to be slow, implying that transport is dominated by diffusion. In other words, the P{\'{e}}clet number $\mathrm{Pe}=VL/\mathcal{D}_m$ is assumed to be small, where $V$ is the characteristic fluid velocity, $L$ the characteristic streamwise length scale, and $\mathcal{D}_m$ the molecular diffusivity. However, it is straightforward to show that $\mathrm{Pe}$ can be order 1 or larger in a tissue or gel subject to fast ($0.1$--$1$~Hz) and large ($10$--$20$\%) deformations (see table II), suggesting that dispersion may be important or even dominant in some scenarios \cite{DELGADO2007}. Indeed, many studies highlight a transition from diffusion-dominated to advection-dominated transport without acknowledging the potential role of dispersion \cite{gardiner2007, Urciuolo2008, DiDomenico2017, Vaughan2013}. With an analogous argument, \citet{Davit2013} illustrated the importance of including dispersion in models for solute transport in biofilms. The second typical reason for neglecting dispersion in tissues and gels is the assumption that the longitudinal and transverse dispersivities themselves are negligible. This expectation is a result of physical insight derived from transport in granular materials, where the dispersivity is typically taken to be proportional to the pore size \cite{Saffman1959, OSWALD2004, Kree2017, Liang2018}. Indeed, the typical pore size is $\sim10$ nm in polymeric gels and in the extra-cellular matrix of tissues (\textit{e.g.}, around 6~nm in cartilage \cite{mow1984}) and can therefore be similar to (or smaller than) the size of large solute molecules \cite{maroudas1970distribution, DiDomenico2018}, originating solid-solute friction \cite{Ateshian2011,yao2007convection}. However, tissues and scaffolds are heterogeneous and multiscale materials; the presence of other components, such as collagen fibres, originates a ``mesoscale'' of larger pores (\textit{e.g.}, 100--150~nm in cartilage \cite{maroudas1975biophysical, levick1987, Federico2008}), where even larger solute molecules can pass \cite{didomenico2017antibody, DiDomenico2018} and where dispersion is likely to play a much larger role. The same is true for double-porosity scaffolds and gels, where additional channels and/or macroscopic pores are included to enhance fluid flow throughout the scaffold depth \cite{denbuijs2010, Lee2015, Mesallati2013}. 
As a further counter-argument, we hypothesise that, even in pores that are small compared to the solute molecules, the irrelevance of pore-scale velocity gradients does not exclude velocity variations and streamline alterations in the overall network, which could originate longitudinal and transverse dispersion. This hypothesis is consistent with the quantification of tortuosity in several soft tissues \cite{maroudas1970distribution, hrabe2004, zhang2005transport}.

Thus, the impact of periodic loading on solute transport in soft porous media has been addressed with various approaches and assumptions across a variety of specific applications in soils, tissues, hydrogels, and scaffolds. However, no single study has yet provided a comprehensive understanding across a wide range of loading frequencies and amplitudes. Moreover, the impact of hydrodynamic dispersion remains relatively unexplored and therefore poorly understood, particularly in the context of biological and biomedical applications. Here, we study the transport and mixing of solutes due to arbitrarily large, periodic deformations of a soft porous material. For the flow and deformation, we adopt a one-dimensional, large-deformation poroelasticity model that includes rigorous nonlinear kinematics, deformation-dependent permeability, and Hencky elasticity for the solid skeleton. In a companion study, we used this model to explore the poromechanics of large-amplitude periodic loading~\citep{Fiori2023}. Here, we additionally consider solute transport due to advection, diffusion, and dispersion. We first study the separate roles of advection, diffusion, and dispersion during one loading cycle. We then consider the impact of the transport and loading parameters on transport and mixing over longer time periods and/or larger numbers of loading cycles. We report the impact of a wide range of loading amplitudes and periods on each transport mechanism and observe how transport depends on the poromechanical response through its impact on local fluid flow. When dispersion is negligible, we show that diffusion is insensitive to loading period but slightly suppressed by increased loading amplitude. With dispersion, larger amplitudes always boost solute spreading;  however, progressively shorter periods impact transport and mixing in more complex ways: fast loading promotes spreading by inducing large fluid velocities, but very fast loading hinders spreading by progressively localising the flow and deformation. We show that the competition between these two effects results in maximum solute transport and mixing for intermediate loading periods.  

\section{Theoretical Model}

Our model combines large-deformation poroelasticity with solute transport. The coupling between periodic deformations and solute movement occurs primarily via the fluid flow, which is originated by the former and responsible for the latter. 

\subsection{Model Problem}

\begin{figure}
    \centering
    \includegraphics[width=0.5\textwidth]{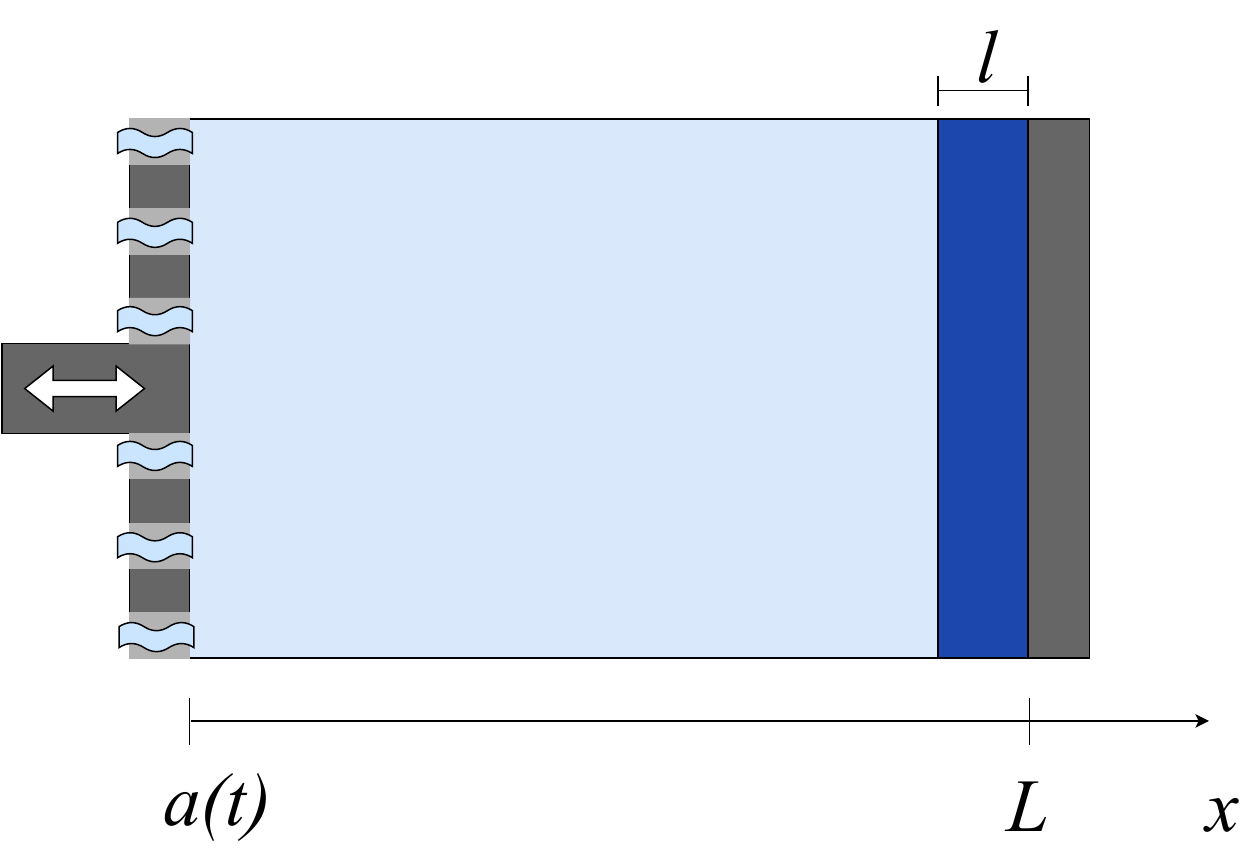}
    \caption{ We consider a 1D sample of soft porous material of relaxed length $L$, subject to a periodic, displacement-driven loading at its left boundary (white arrows). The left boundary is permeable, thus allowing fluid flow in or out (pale blue squiggles) to accommodate the loading. The right boundary is fixed and impermeable. The solute is initially localised against the right boundary in a strip of width $l$ (dark blue). \label{system} }
\end{figure}

We consider a one-dimensional sample of soft porous material of relaxed length $L$ and relaxed porosity (fluid fraction) $\phi_{f,0}$. The left boundary of the material (at $x=a(t))$ is moving and permeable, whereas the right boundary (at $x=L$) is fixed and impermeable. The position of the left boundary, $a(t)$, is imposed to be:
\begin{equation}\label{eq:a(t)}
    a(t)= \frac{A}{2} \left[1-\cos\left(\frac{2\pi t}{T}\right)\right],
\end{equation}
where $A$ and $T$ are the amplitude and period of loading, respectively. We consider imposed deformations ranging from small to large macroscopic strains ($-0.4\%$ to $-20\%$ or $0.004\leq{}A/L\leq{}0.2$). We take the fluid and solid to be individually incompressible, such that changes in bulk volume correspond directly to the movement of fluid into and out of the pore space. We presented and analysed the poromechanics of this scenario in detail in a companion study \cite{Fiori2023}. We now introduce a strip of passive solute of initial width $l$ located at the right boundary and we study the impact of this periodic, displacement-driven deformation on the evolution of the solute distribution (figure~\ref{system}).

\subsection{Kinematics}\label{kinematics}

We consider an Eulerian reference frame, in which the solid displacement is $\mathbf{u_{s}}=\mathbf{x}-\mathbf{X}(\mathbf{x},t)$, with $\mathbf{X}(\mathbf{x},t)$ the reference position of the solid material point that at time $t$ occupies position $\mathbf{x}$. We choose our reference configuration to be the relaxed configuration, such that $\mathbf{X}(\mathbf{x},0)=\mathbf{x}$ and $\mathbf{u_{s}}(\mathbf{x},0)=0$. The true volume fractions of fluid and solid are $\phi_f$ and $\phi_{s}$, respectively, where $\phi_{f}+\phi_{s}=1$. In this uniaxial setting, the solid displacement and the solid and fluid velocities are one-dimensional and given by 
\begin{equation}\label{eq:displacements}
    \mathbf{u_{s}}=u_{s}(x,t)\mathbf{\hat{e}_{x}}, \;
    \mathbf{v_{s}} = v_{s}(x,t)\mathbf{\hat{e}_{x}}, \; 
    \mathbf{v_{f}} = v_{f}(x,t)\mathbf{\hat{e}_{x}}, 
\end{equation}
where $\mathbf{v_f}$ and $\mathbf{v_s}$ are the fluid and solid velocities, respectively, $u_s$, $v_s$, and $v_f$ are the $x$-components of these fields, and $\mathbf{\hat{e}_{x}}$ is the unit vector in the $x$-direction. The local current volume per unit reference volume is measured by the Jacobian determinant, which in this uniaxial setting is given by $J= (1-\partial{u_s}/\partial{x})^{-1}$. For incompressible constituents and uniform initial porosity $\phi_{f,0}$, the local change in volume relates to the change in porosity as
\begin{equation}\label{porosity-def}
    J(x,t) = \frac{1-\phi_{f,0}}{1-\phi_f} \quad\to\quad \frac{\partial{u_s}}{\partial{x}} =\frac{\phi_f-\phi_{f,0}}{1-\phi_{f,0}}.
\end{equation}
Continuity can be written
\begin{equation}\label{continuity}
    \frac{\partial{\phi_f}}{\partial{t}} +\frac{\partial}{\partial{x}}(\phi_f v_f) = 0 \quad\mathrm{and}\quad \frac{\partial{\phi_f}}{\partial{t}} -\frac{\partial}{\partial{x}}{[(1-\phi_f)v_s]} = 0,
\end{equation}
which together imply that the total flux $q=\phi_fv_f+(1-\phi_f)v_s$ is uniform in space, $\partial{q}/\partial{x}=0$.

\subsection{Fluid flow}

We assume that the fluid flows relative to the solid according to Darcy's law: 
\begin{equation}\label{darcylaw}
    \phi_ f (v_{f}-v_{s}) = - \frac{k(\phi_f)}{\mu}\frac{\partial p}{\partial x},
\end{equation}
where $k(\phi_f)$ is the permeability of the solid skeleton, $\mu$ is the dynamic viscosity of the fluid, and $p$ is the fluid (pore) pressure, and where we have neglected gravity.
As in Ref.~\cite{Fiori2023}, we take the permeability to be deformation-dependent according to a normalised Kozeny-Carman relation, $k(\phi_{f}) = k_0 \frac{(1-\phi _{f,0})^2}{\phi _{f,0}^3} \frac{\phi _{f}^3}{(1-\phi _{f})^2}$,
where $k_0\equiv{}k(\phi_{f,0})$ is the permeability of the initial state. We discuss this choice in detail in Ref.~\cite{Fiori2023}. 

Combining equations \eqref{continuity} and \eqref{darcylaw}, we arrive at the nonlinear flow equations: 
\begin{equation}\label{conservation-q}
    \frac{\partial{\phi_f}}{\partial{t}} +\frac{\partial}{\partial{x}}\bigg[{\phi_f q}-(1-\phi_f) \frac{k(\phi_f)}{\mu}\frac{\partial{p}}{\partial{x}}\bigg]=0 \quad\mathrm{and}\quad \frac{\partial{q}}{\partial{x}}=0,
\end{equation}
where the total flux $q$ is again
\begin{equation}\label{q-vf-vs}
    q\equiv \phi_f v_f + (1-\phi_f)v_s
\end{equation}
and the fluid and solid velocities are given by
\begin{equation}
    v_f=q-\frac{(1-\phi_f)}{\phi_f}\frac{k(\phi_f)}{\mu}\frac{\partial p}{\partial x}   
    \quad\mathrm{and}\quad
    v_s=q+\frac{k(\phi_f)}{\mu}\frac{\partial p}{\partial x}.   
\end{equation}
Note that the fluid flux is
\begin{equation}\label{flux-qf}
    q_f=\phi_f v_f.
\end{equation}

\subsection{Mechanical equilibrium and elasticity law}\label{elasticity_skeleton}

Neglecting inertia, gravity and other body forces, mechanical equilibrium can be expressed as  ${\boldsymbol{\nabla}\cdot{\bm{\sigma}}}= {\boldsymbol{\nabla}\cdot{\bm{\sigma'}}}-\boldsymbol{\nabla} p= 0$, where $\bm{\sigma}$ is the true Cauchy total stress, decomposed into contributions from the fluid pressure $p$ and from Terzaghi's effective stress $\bm{\sigma'}$. In 1D, mechanical equilibrium reads 
\begin{equation}\label{sigma-p}
    \frac{\partial \sigma'}{\partial x}=\frac{\partial p}{\partial x},
\end{equation}
where $\sigma^\prime$ is the $xx$ component of $\boldsymbol{\sigma}^\prime$.

We take the solid skeleton to be elastic, with no viscous or dissipative behaviours. Since any elasticity law can be written in the form $\sigma^\prime=\sigma^\prime(\phi_f)$ for a uniaxial deformation, this problem can be described by a  nonlinear advection-diffusion equation:
\begin{equation}\label{eq:conservation-q-Df}
    \frac{\partial{\phi_f}}{\partial{t}} +\frac{\partial}{\partial{x}}\bigg[{\phi_f q}-D_f(\phi_f)\frac{\partial{\phi_f}}{\partial{x}}\bigg]=0 \quad\mathrm{and}\quad \frac{\partial{q}}{\partial{x}}=0,
\end{equation}
where the nonlinear composite constitutive function
\begin{equation}\label{eq:Df}
    D_f(\phi_f)=(1-\phi_f)\frac{k(\phi_f)}{\mu}\frac{\mathrm{d}\sigma^\prime}{\mathrm{d}\phi_f}
\end{equation}
is the poroelastic diffusivity. Note that a very similar model is used for the solidification of colloidal suspensions in applications such as filtration and sedimentation, for which the poroelastic diffusivity $D_f(\phi_f)$ (\textit{i.e.}, the ``solids diffusivity'') is characterised as a composite material property~\citep[\textit{e.g.},][]{Davis1989, peppin2006, Style2011, bouchaudy-softmatter-2019, Worster2021}.

We use Hencky hyperelasticity~\cite{hencky} as a simple, large-deformation model that captures kinematic nonlinearity. For a uniaxial deformation, the relevant component of the effective stress is then
\begin{equation}\label{sigma-hencky}
    \sigma^\prime= \mathcal{M}\frac{\ln(J)}{J}=\mathcal{M} \bigg(\frac{1-\phi_f}{1-\phi_{f,0}}\bigg) \ln\bigg(\frac{1-\phi_{f,0}}{1-\phi_f}\bigg),
\end{equation}
where $\mathcal{M}$ is the $p$-wave or oedometric modulus~\citep{Macminn2016}. Note that, for these constitutive choices of Kozeny-Carman permeability and Hencky elasticity, the permeability at the left boundary can vanish for sufficiently large $A$ and/or small $T$, because the poroelastic diffusivity remains finite rather than diverging as $\phi_f\to0$~\citep{hewitt2016}. We motivate and discuss our constitutive choices and explore the poroelastic diffusivity in more detail in Ref.~\cite{Fiori2023}.

With appropriate initial conditions, boundary conditions, and the normalised Kozeny-Carman permeability law, equations~\eqref{eq:conservation-q-Df}, \eqref{eq:Df}, and \eqref{sigma-hencky} comprise a closed model for the evolution of the porosity.

\subsection{Solute transport}

We now consider the transport of solute. We denote the true local solute concentration in the fluid phase by $c$ (amount of solute per unit current fluid volume). We take the solute to be passive and charge-neutral, with no chemical or other interaction with the solid or fluid phases, so that neither the fluid properties nor the solid properties depend on $c$. The flow and mechanics above are then independent of the transport problem. 

In 1D, it is well known that conservation of mass for a passive solute can be written
\begin{equation}\label{conservation-c}
    \frac{\partial }{\partial t} (\phi_f c) +\frac{\partial}{\partial x}\bigg[{\phi_f c v_f }- \phi_f \mathcal{D}   \frac{\partial c}{\partial x}\bigg]=0.
\end{equation}
The first term in the square brackets is the Darcy-scale solute flux due to advection, which occurs here entirely in response to the deformation.
The second term in the square brackets combines molecular diffusion and hydrodynamic dispersion, thus taking the latter to be a Fickian process  \citep[\textit{e.g.},][]{Scheidegger1961}. The latter term is multiplied by the porosity $\phi_f$ since solute movements only occur in the fluid phase. The coefficient $\mathcal{D}$ can be written
\begin{equation}\label{Diff}
    \mathcal{D}   =\mathcal{D}_m+\mathcal{D}_h,
\end{equation}
where $\mathcal{D}_m$ and $\mathcal{D}_h$ are the coefficients of molecular diffusion and hydrodynamic dispersion, respectively. Dispersion, in which pore-scale velocity gradients and the tortuosity of the pore space lead to macroscopic spreading of solute, depends sensitively on flow conditions and the details of the pore structure in ways that are not yet fully understood, even for rigid porous materials~\citep{Dentz2018, Dentz2023}. The most widely used model for the macroscopic dispersive flux is Fickian, as above, with a velocity-dependent dispersion coefficient given in 1D by
\begin{equation}\label{Disp}
    \mathcal{D}_h=\alpha |v_f-v_s|,
\end{equation}
where $\alpha$ is the longitudinal dispersivity~\citep{Scheidegger1961, brenner1993, whitaker1999method, gelhar1993stochastic}. Note that the dispersive flux is proportional to $|v_f-v_s|$, unlike the advective flux, because dispersion is driven by  flow of fluid \emph{through} the pore structure (\textit{i.e.}, $v_f=v_s\neq 0$ would lead to advection but no dispersion). Note also that, unlike the advective flux, the diffusive and dispersive fluxes are independent of the direction of the fluid flow.

The dispersivity $\alpha$ is typically taken to be a constant material property for a given pore structure. In a deforming porous material, and particularly for moderate to large deformations, it is likely that $\alpha$ should be deformation-dependent to account for the evolving pore structure. For example, particle-particle interactions and rearrangements are known to drive enhanced dispersion in dense suspensions~\cite{Souzy2016, souzy2017} and compaction has been shown to have a nontrivial impact on dispersion in bead packs and packed beds~\cite{charlaix-physfluids-1987, oestergren-chemengj-2000, liu-jfm-2024}. We expect similar but even larger effects in poroelastic materials under large deformations, which may ultimately require novel dispersion models, but these phenomena are beyond the scope of the present study. Here, we take $\alpha$ to be a constant for simplicity.

\subsection{Initial and boundary conditions}

We next specify initial and boundary conditions for the solid, the fluid, and the solute. Recall that the left and right boundaries of the solid are at $x=a(t)$ and $x=L$, respectively. 

\subsubsection{Initial conditions}

Equation~\eqref{eq:a(t)} implies that $a(0)=0$, and thus that the initial porosity is uniform and equal to the relaxed porosity
\begin{equation}
   \phi_f(x,0)= \phi_{f,0} \textrm{ and } u_s(x,0)=0.
\end{equation}
We take the solute to be initially localised against the right boundary in a strip of width $l$ and concentration $c_0$, such that
\begin{equation}
    c(x,0)=\frac{c_0}{2} \left\{ \tanh{[s(x-L+l)}]+1 \right\},
\end{equation}
where $s$ is a steepness parameter. 

\subsubsection{Left boundary}

For $t>0$, we apply a displacement-controlled loading at the left boundary according to equation~\eqref{eq:a(t)}. We take this moving boundary to be fluid- and solute-permeable. The associated boundary conditions are
\begin{equation}\label{bc-left}
    u_s(a,t)=a(t), \quad v_s(a,t)=\frac{\mathrm{d}a}{\mathrm{d}t}\quad\mathrm{and}\quad 
    p(a,t)=0.
\end{equation}  
We take the fluid outside the domain to be ``clean'', such that 
\begin{equation}
    c(a,t)=0.
\end{equation}

\subsubsection{Right boundary}

We take the right boundary to be fixed and impermeable, such that
\begin{equation}\label{bc-right}
    u_s(L,t)=v_s(L,t)=v_f(L,t)=0 \quad \mathrm{and} \quad 
    \frac{\partial c}{\partial x}\bigg|_{x=L}=0.
\end{equation}
Equation~\eqref{bc-right} and the requirement that $q$ be uniform in space imply that  there can be no net flow from left to right in our problem, $q\equiv0$. Equation~\eqref{q-vf-vs} then implies that the fluid and the solid always locally move in opposite directions,
\begin{equation}\label{vf-vs}
    v_f= -\frac{(1-\phi_f)}{\phi_f} v_s.
\end{equation}

\subsection{Scaling and summary}\label{scaling}

As in Ref.~\cite{Fiori2023}, we apply the following non-dimensionalization to the poromechanical model
\begin{equation}
\begin{split}
    \tilde{x}=\frac{x}{L},\;
    \tilde{u}_s=\frac{u_s}{L},\;
    \tilde{t}=\frac{t}{T_{\mathrm pe}},\;
    \tilde{\sigma}^\prime=\frac{\sigma'}{\mathcal{M}},\;
    \tilde{p}=\frac{p}{\mathcal{M}},\; \tilde{k}=\frac{k(\phi)}{k_0},\; \tilde{v}_f=\frac{v_f}{L/T_{\mathrm pe}}, \; \tilde{v}_s=\frac{v_s}{L/T_{\mathrm pe}},
\end{split}
\end{equation}
where $T_{\mathrm{pe}}=L^2/D_{f,0}=\mu{}L^2/(k_0\mathcal{M})$ is the classical poroelastic timescale for the relaxation of pressure over a distance $L$ and $D_{f,0}=k_0\mathcal{M}/\mu$ is the constant linear-poroelastic diffusivity.

We then scale quantities related to solute transport as
\begin{equation}
    \tilde{c}=\frac{c}{c_0},  \; 
    \tilde{l}=\frac{l}{L}, \; \tilde{\alpha}=\frac{\alpha}{L}.
\end{equation}

Taking $q\equiv{}0$, as noted above, the full problem can then be rewritten in dimensionless form as
\begin{equation}\label{conservation-q-scaled}
     \frac{\partial{\phi_f}}{\partial{\tilde{t}}} -\frac{\partial}{\partial{\tilde{x}}}\bigg[\tilde{D}_f(\phi_f)\frac{\partial{\phi_f}}{\partial{\tilde{x}}}\bigg]=0,
\end{equation}
where
\begin{equation}
    \tilde{D}_f=\frac{D_f}{D_{f,0}}=(1-\phi_f)\tilde{k}(\phi_f)\frac{\mathrm{d}\tilde{\sigma}^\prime}{\mathrm{d}\phi_f},
\end{equation}
and
\begin{equation}\label{conservation-c-scaled}
    \frac{\partial }{\partial \tilde{t}}(\phi_f \tilde{c})  +\frac{\partial}{\partial \tilde{x}}\bigg[{\phi_f \tilde{c} \tilde{v}_f }- \phi_f \tilde{\mathcal{D}}  \frac{\partial \tilde{c}}{\partial \tilde{x}}\bigg]=0.
\end{equation}
The dimensionless coefficient of diffusion/dispersion $\Tilde{\mathcal{D}}$ is
\begin{equation}
    \tilde{\mathcal{D}}=\frac{\mathcal{D}}{\mathcal{D}_m}=\mathrm{Pe}^{-1}+\tilde{\alpha} | \tilde{v}_f-\tilde{v}_s|,  
\end{equation}
where $\mathrm{Pe}=\frac{L^2/T_{\mathrm pe}}{\mathcal{D}_m}=\frac{k_0 \mathcal{M}}{\mu \mathcal{D}_m}$ is the P{\'{e}}clet number, which measures the importance of poroelastic-relaxation-driven advection relative to molecular diffusion.

The initial conditions are
\begin{equation}
   \tilde{a}(0)=0, \;
   \phi_f(\tilde{x},0)= \phi_{f,0},
\end{equation}
and 
\begin{equation}
    \tilde{c}(\tilde{x},0)=\frac{1}{2} \{ \tanh{[\tilde{s}(\tilde{x}-1+\tilde{l})}]+1\},
\end{equation}
where we take $\tilde{s}=sL=60$. The boundary conditions are
\begin{equation}\label{leftB}
    \tilde{u}_s(\tilde{a},\tilde{t})=\tilde{a}(\tilde{t})= \frac{\tilde{A}}{2} \Bigg[1-\cos\left(\frac{2\pi\tilde{t}}{\tilde{T}}\right)\Bigg] \,,\,\,\tilde{v}_s(\tilde{a},\tilde{t})=\frac{\mathrm{d} \tilde{a}}{\mathrm{d}\tilde{t}}\,,\,\ \quad \tilde{p}(\tilde{a},\tilde{t})=0, \; \mathrm{and} \quad \tilde{c}(\tilde{a},\tilde{t})=0
\end{equation}
and 
\begin{equation}
    \tilde{u}_s(1,\tilde{t}) =\tilde{v}_s(1,\tilde{t}) =\tilde{v}_f(1,\tilde{t})=0 \quad\mathrm{and}\quad \frac{\partial\tilde{c}}{\partial \tilde{x}}\bigg|_{\tilde{x}=1}=0,
\end{equation}
where $\tilde{A}=A/L$ and $\tilde{T}=T/T_{\mathrm{pe}}$. 

As shown in Ref.~\cite{Fiori2023}, the forcing considered here will drive a typical solid velocity of size $v_s^*=2A/T$ and thus a typical fluid velocity of size $v_f^*=\left(\frac{1-\phi_{f,0}}{\phi_{f,0}}\right)(2A/T)$. The characteristic advection time $T_\mathrm{adv}$ and diffusion time $T_\mathrm{diff}$ are then
\begin{equation}
    T_\mathrm{adv}=\frac{L}{v_f^*} =\frac{LT\phi_{f,0}}{2A(1-\phi_{f,0})} \quad\to\quad \tilde{T}_\mathrm{adv} =\frac{T_\mathrm{adv}}{ T_\mathrm{pe}} =\frac{\tilde{T}\phi_{f,0}}{2\tilde{A}(1-\phi_{f,0})} \propto \frac{\tilde{T}}{\tilde{A}}
\end{equation}
and
\begin{equation} 
    T_\mathrm{diff}=\frac{L^2}{\mathcal{D}_m} \quad\to\quad \tilde{T}_\mathrm{diff}=\frac{{T}_\mathrm{diff}}{T_\mathrm{pe}} = \frac{D_{f,0}}{\mathcal{D}_m}=\mathrm{Pe}
\end{equation}
Recall that the P{\'{e}}clet number --- as defined above --- quantifies the rate of advection due to poromechanical relaxation relative to the rate of molecular diffusion. The characteristic times above suggest that the balance between loading-driven advection and molecular diffusion is better measured by an effective P{\'{e}}clet number $\mathrm{Pe}_\mathrm{eff}$,
\begin{equation}
    \mathrm{Pe_\mathrm{eff}} =\mathrm{Pe}\,\frac{\tilde{A}}{\tilde{T}} \propto \frac{\tilde{T}_\mathrm{diff}}{\tilde{T}_\mathrm{adv}}.
\end{equation}
In our results below, we explore a wide range of $\mathrm{Pe_\mathrm{eff}}$: from $\sim$1 to $\sim$$10^5$. We show in Table~\ref{table-peclet} that this range is biologically relevant.

The above model describes uniaxial flow, mechanics, and solute transport in a poroelastic material subject to periodic deformations. The kinematics are rigorous and thus nonlinear, the elasticity law is Hencky elasticity, and the permeability law is the normalised Kozeny-Carman formula. Solute transport occurs via advection, molecular diffusion, and hydrodynamic dispersion. We solve this system numerically in MATLAB using compact finite differences in space and an implicit Runge-Kutta method in time, as described in more detail in Appendix~\ref{numerical-method}. We provide an example code in \citet{fiori-jfm-2025_code}. Below, we consider only dimensionless quantities, dropping the tildes for convenience.

\section{Solute transport and mixing}

\subsection{Quantification of solute transport and mixing}\label{Variables}

\begin{figure}
    \centering
    \includegraphics[height=8cm]{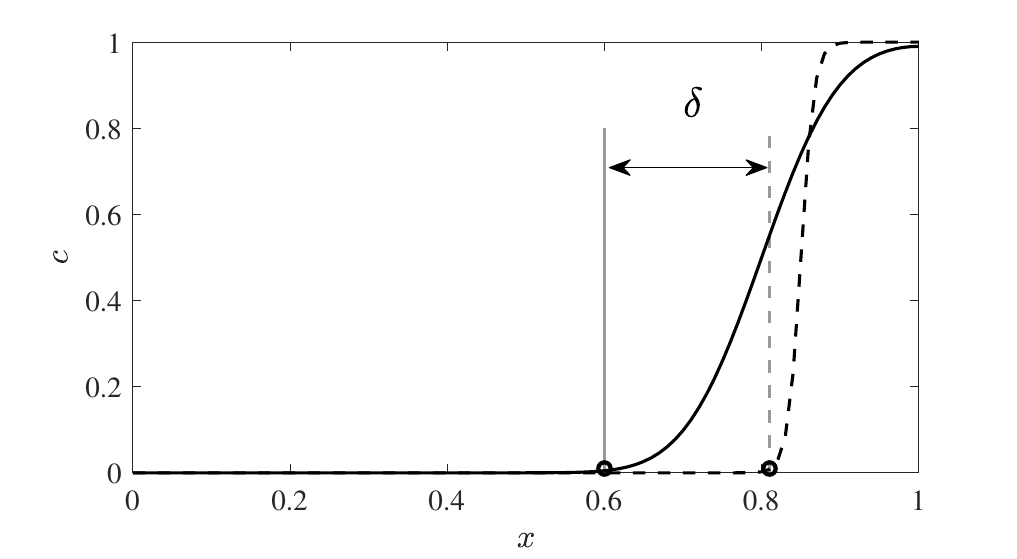}
    \caption{ Schematic representation of the travel distance or mixing length $\delta$, which measures the distance traveled by the left edge of the concentration profile during the time $t$. For solute initially localised in a finite strip at the right, we calculate $\delta(t)$ by choosing a small threshold concentration $c_{\delta}$ and then finding the leftmost position $x_{\delta}(t)$ where that concentration occurs. Then, $\delta(t) = |x_{\delta}(t) - x_{\delta}(0) |$ \citep[see, \textit{e.g.},][]{tan1988simulation, mishra2008differences}. Here, we show $c(x,0)$ (dashed curve), $c(x,t)$ (solid curve), and the corresponding $\delta(t)$. The value of $c_\delta$ is arbitrary and should have no qualitative impact on the results. In the results shown below, we take $c_{\delta}= 0.01$. \label{fig:ld} }
\end{figure}

We begin with some qualitative examples that illustrate the impact of deformation on each transport mechanism individually. We also assess solute transport and mixing quantitatively via two metrics:  
\begin{enumerate}
    \item The travel distance or mixing length $\delta$ measures the distance travelled by the left edge of the concentration profile (figure~\ref{fig:ld}). The travel distance can range from $0$ to $1-l$, but it becomes less meaningful as it approaches $1-l-A$, by which point the concentration profile interacts strongly with the left boundary.
    \item The degree of mixing $\chi$ measures the degree to which the initial concentration profile has homogenised, and is closely related to the variance of the concentration distribution. We express the degree of mixing in terms of the variance of the concentration distribution by generalising the standard definition \citep[see, \textit{e.g.},][]{Danckwerts1952, jha2011quantifying} to account for a porosity field that varies in space. Considering the fluid-volume-weighted average $\langle\ast\rangle_{f}$, defined as 
    \begin{align}\label{eq:spatial_average}
        \langle\ast\rangle_f=\frac{\int_a^1\,\phi_f\ast\,\mathrm{d}x}{\int_a^1\,\phi_f\,\mathrm{d}x},
    \end{align}
    the variance of the concentration distribution is then
    \begin{equation}\label{eq:variance}
        \sigma^2(t) = \langle{}c^2\rangle{}_{f} - \langle{}c\rangle{}_{f}^2 
    \end{equation}
    and the degree of mixing is 
    \begin{equation}\label{eq:chi}
        \chi(t) = 1 - \frac{\sigma^2(t)}{\sigma_{\rm max}^2},
    \end{equation}
    where $\sigma_{\rm max}^2=\sigma^2(t=0)$ in this case. Note that $\chi$ can range from $0$ to $1$, where the former corresponds to no mixing (\textit{i.e.}, the initial state by definition) and the latter is characteristic of a completely mixed configuration (\textit{i.e.}, spatially uniform concentration).
\end{enumerate}

\subsection{Baseline values}\label{baseline}

For a given total loading time, $\delta$ and $\chi$ depend on the transport parameters $\mathrm{Pe}^{-1}$ and $\alpha$; the loading parameters $A$ and $T$; the initial porosity $\phi_{f,0}$; and the initial width of the solute strip $l$. We choose a baseline value for each parameter (table~\ref{tab:quantitative}). We use these baseline values in all of the results presented below, except where explicitly noted otherwise. We explore the impact of individually changing $\mathrm{Pe}^{-1}$ and $\alpha$ in section \ref{mat-parameters},  $A$ and $T$ in section \ref{ST-different-regimes}, and $\phi_{f,0}$ and $l$ in Appendix \ref{appendix-phi0-l}.

We choose a baseline amplitude $A=0.1$, corresponding to moderately large deformations. We choose a baseline period $T=6\pi$, which, following our companion study (Ref. \cite{Fiori2023}), ensures that the poromechanics are quasi-static (\textit{i.e.},  ``slow-loading''; see the first part of section \ref{ST-different-regimes}). The fluid flux $q_f$ and the relative velocity $|v_f-v_s|$ for this baseline case are shown in figure \ref{fig:flux-T}d and \ref{fig:flux-T}h, respectively. The baseline values of $\mathrm{Pe}^{-1}$ and $\alpha$ are in the range of those proposed by \citet{Sengers2004} for cartilage constructs, with the specific values chosen to ensure that diffusion dominates over dispersion for the slowest period considered in this study (see Appendix \ref{appendix-dispersive}). The baseline value for $\phi_{f,0}$ is representative of hydrogels or soft biological tissues, whereas $l$ is arbitrarily chosen to be a small fraction of the domain length.

\begin{longtblr}[
    caption={Baseline parameter values.},
    label = {tab:quantitative},
    entry = none,
    ]{
    width = \linewidth,
    colspec = {Q[185]Q[77]Q[83]Q[235]Q[121]Q[142]Q[90]},
    cells = {c},
    vline{2-7} = {-}{},
    hline{2} = {-}{},
    }
    Parameter       & $A$   & $T$    & $\mathrm{Pe}^{-1}$ & $\alpha$       & $\phi_{f,0}$ & $l$    \\
    Baseline~value~ & $0.1$ & $6\pi$ & $3\times10^{-5}$   & $0.01$ & $0.75$          & $0.05$
\end{longtblr}

\subsection{Qualitative impacts of periodic loading on solute transport}\label{isolation-mechanisms}

\begin{figure}
    \centering
    \includegraphics[width=0.95\textwidth]{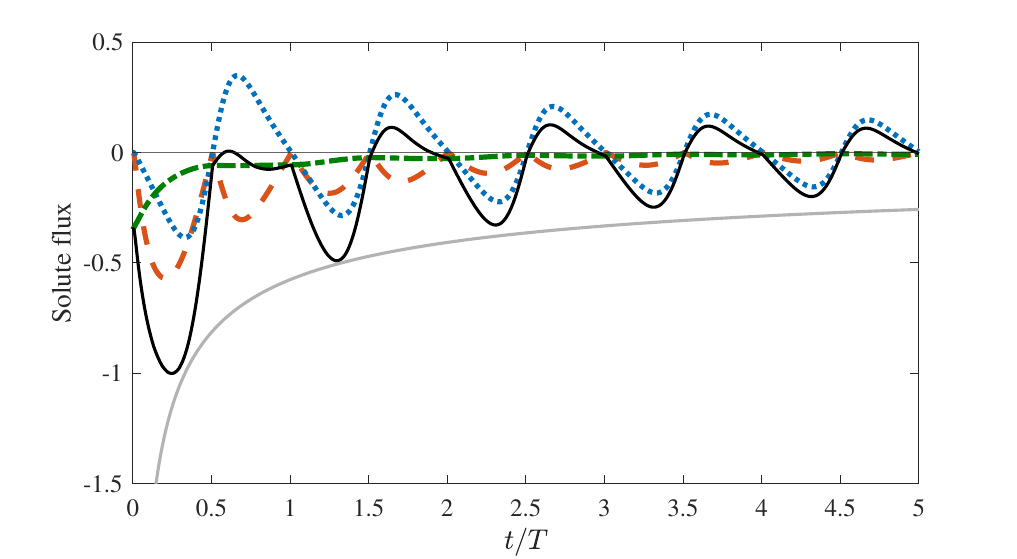}
    \caption{ Evolution of the solute flux across $x=1-l$ during 5 loading cycles. We show the total flux of solute (solid black) and the separate contributions of advection (dotted blue), molecular diffusion (dash-dotted green), and hydrodynamic dispersion (dashed red) for $A=0.4, \alpha=0.025$. Note that $A$ and $\alpha$ are higher than the baseline values to better illustrate the roles of advection and dispersion. The solid grey envelope is proportional to $t^{-\frac{1}{2}}$. \label{fig:fluxes} }
\end{figure}

\begin{figure}
    \centering
    \includegraphics[width=0.475\textwidth,trim= 4cm 0.6cm 4cm 1.cm]{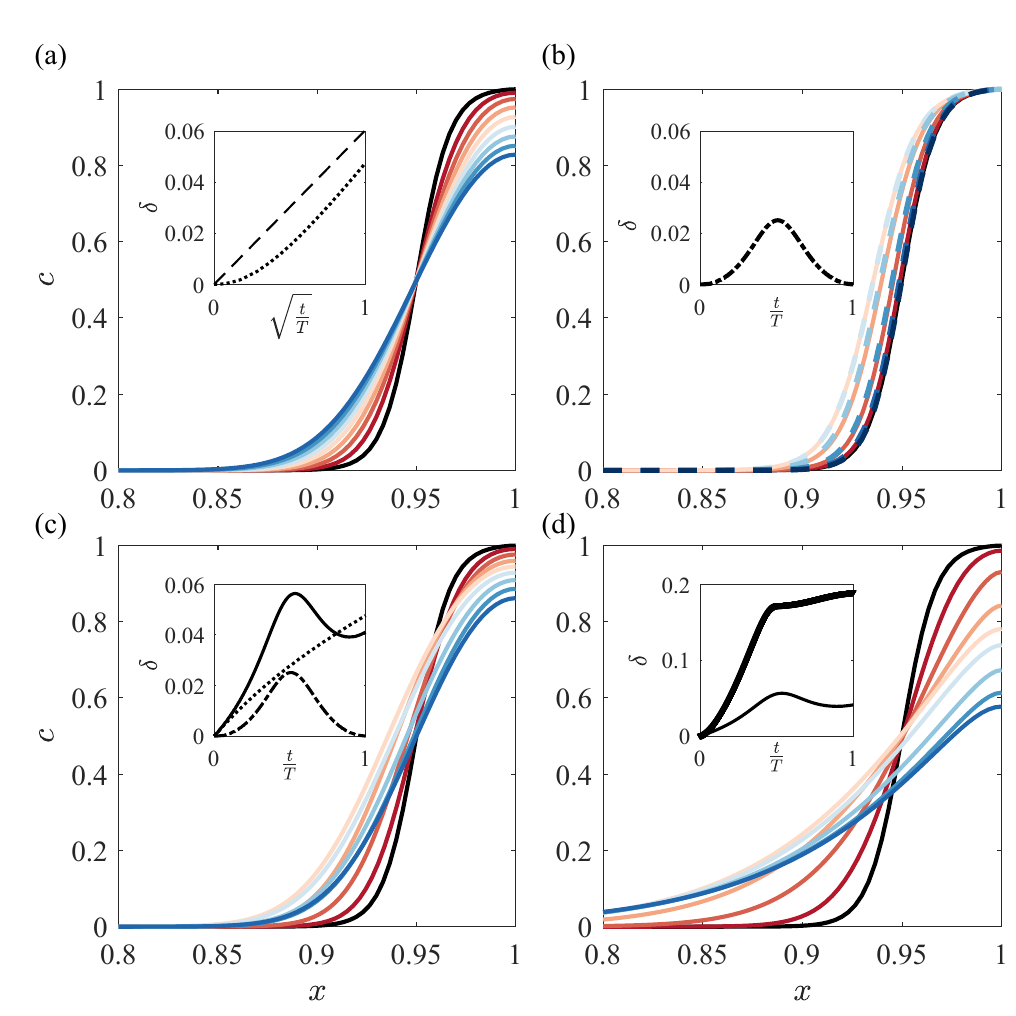}
    \caption{ Evolution of the concentration profile during one cycle (red to blue through white) for four cases: (a) diffusion only ($A=\alpha=0, \mathrm{Pe}^{-1}=3\times10^{-5}$); (b) advection only  ($A=0.4, \mathrm{Pe}^{-1}=\alpha=0$); (c) advection and diffusion ($A=0.4, \mathrm{Pe}^{-1}=3\times10^{-5}, \alpha=0$); (d) advection, diffusion, and dispersion ($A=0.4, \mathrm{Pe}^{-1}=3\times10^{-5}, \alpha=0.025$). We plot concentration against the spatial coordinate $x$ and split the evolution into two phases, loading ($\dot{a}>0$, first half of the cycle, dark to light red) and unloading ($\dot{a}<0$, second half, light to dark blue). In panel~(b), the unloading curves (dashed) overlap with the loading curves (solid). The initial profile is shown in black. For each case, we also show the evolution of $\delta$ throughout the loading cycle (insets); in all cases, the dotted curves are for diffusion without loading (with the dashed reference line showing linearity with $\sqrt{t/T}$), the dash-dot curves are for advection only, the thin solid curves are for advection and diffusion, and the thick solid curve is for advection, diffusion, and dispersion. Note that $A$ and $\alpha$ are higher than the baseline values to better illustrate the roles of advection and dispersion. \label{fig:conc-profile} }
\end{figure}

We begin by isolating and comparing the solute transport mechanisms.
To illustrate the contribution of each mechanism, we consider the time evolution of their separate contributions to the total solute flux at $x=1-l$, which is the initial left edge of the concentration profile, during five loading cycles (figure~\ref{fig:fluxes}). The individual contribution of the advective, diffusive, and dispersive solute fluxes are $q_\mathrm{adv}=\phi_f v_f c  $, $q_\mathrm{diff}=\phi_f \mathrm{Pe}^{-1} (\partial c/\partial x)$, and $q_\mathrm{disp}=\phi_f \alpha|v_f-v_s| (\partial c/\partial x)$, respectively. 
During the loading half of each cycle ($\dot{a}>0$), all three fluxes are negative, implying that all three mechanisms drive solute to the left. During the unloading half of each cycle ($\dot{a}<0$), however, the flow changes direction and the advective flux changes sign (now positive, meaning to the right), whereas the diffusive and dispersive fluxes remain negative (still to the left). The flow and deformation are periodic after an initial transient that decays exponentially (see Ref~\cite{Fiori2023}), in which case the net contribution of advection over one full cycle is zero (see figure~\ref{fig:conc-profile}b). Thus, net  transport at the end of each cycle depends on the cumulative amount of diffusion and dispersion. Diffusion and dispersion are strongest at early times, when the concentration gradient is largest, and decay over time as $t^{-1/2}$. The strengths of diffusion and dispersion are proportional to $\mathrm{Pe}^{-1}$ and $\alpha$, respectively.

We next plot the evolution of the concentration profile during the first cycle (figure~\ref{fig:conc-profile}). We consider four cases: molecular diffusion only, in which $A=0$ (no loading); advection only, in which $\mathrm{Pe}^{-1}=\alpha =0$; advection and molecular diffusion only, in which $\alpha=0$; and the general case, including all three mechanisms. 
For diffusion only (figure~\ref{fig:conc-profile}a) solute spreading is driven exclusively by concentration gradients and the travel distance $\delta$ grows as $\delta \propto \sqrt{t}$ after an initial (slower) phase in which the profile adjusts from its initial condition toward classical self similarity (see Appendix~\ref{anl_dff}). When a deformation is applied (figure~\ref{fig:conc-profile}b-d), four main factors impact the movement of the solute: (i)~the motion of the fluid drives advection; (ii)~the motion of the fluid through the pore space drives dispersion; (iii)~the decrease in porosity weakly hinders diffusion and dispersion since $\phi_f$ is a prefactor in both of those fluxes; and (iv)~the stretched solute profile, with the same quantity of fluid (and solute) now occupying a larger spatial extent, weakens concentration gradients. The latter two effects become increasingly strong during loading and then decreasingly strong during unloading. The fourth mechanism is most obvious in the case of advection only (figure~\ref{fig:conc-profile}b), where the motion of the solute is perfectly reversible and $\delta=0$ at the end of the loading cycle.
The fact that loading weakly suppresses molecular diffusion via the third and fourth mechanisms is apparent in figure~\ref{fig:conc-profile}c, where the final value of $\delta$ is lower for diffusion with loading than for diffusion without loading. When dispersion is included (figure~\ref{fig:conc-profile}d), transport is greatly amplified despite the weak suppression of diffusion.

We next consider several quantitative measures of transport and mixing. 

\subsection{Impact of diffusion and dispersion coefficients on transport and mixing}\label{mat-parameters}

\begin{figure}
    \centering
    \includegraphics[width=1\textwidth]{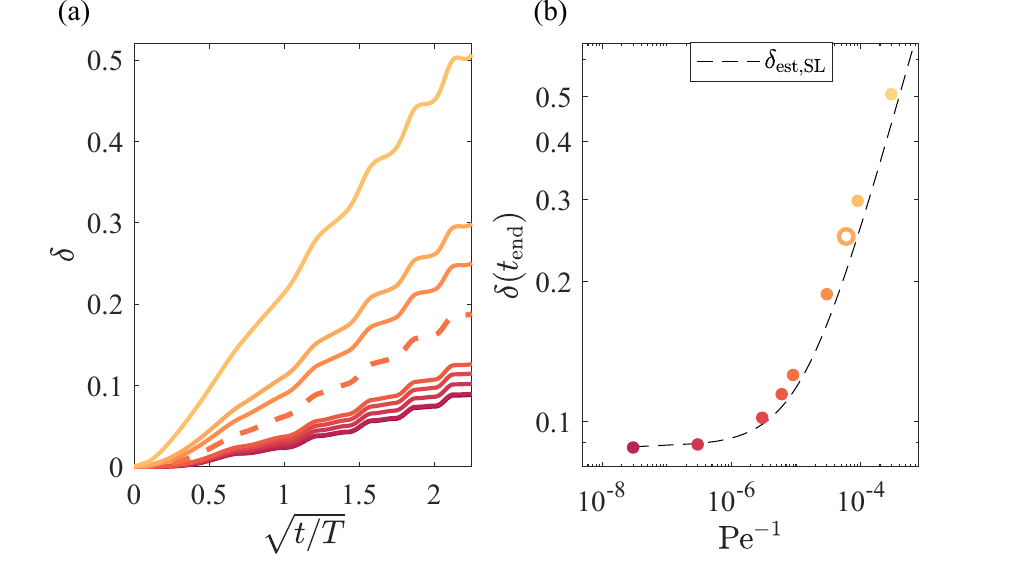}
    \caption{ Impact of $\mathrm{Pe}^{-1}$ on the evolution of $\delta$ over 5 loading cycles. (a) We plot the evolution of $\delta$ with $\sqrt{t}$ for nine different values of $\mathrm{Pe}^{-1}$ $\in [3 \times 10^{-8},3 \times 10^{-4}]$ (dark to light). Note that the curves for the two smallest values of $\mathrm{Pe}^{-1}$ overlap. In each case, delta is roughly linear in $\sqrt{t}$  with a slope that increases monotonically with $\mathrm{Pe}^{-1}$. The dashed curve indicates the baseline value of $\mathrm{Pe}^{-1}$. (b) We plot the final value of $\delta$ at $t=5T$ as function of $\mathrm{Pe}^{-1}$. The open circle indicates the baseline value of $\mathrm{Pe}^{-1}$. The black dashed curve is our estimate $\delta_\mathrm{est, SL}$ from equation~\eqref{delta-estimate-sl}. \label{fig:D0} }
\end{figure}

\begin{figure}
    \centering
    \includegraphics[width=1\textwidth]{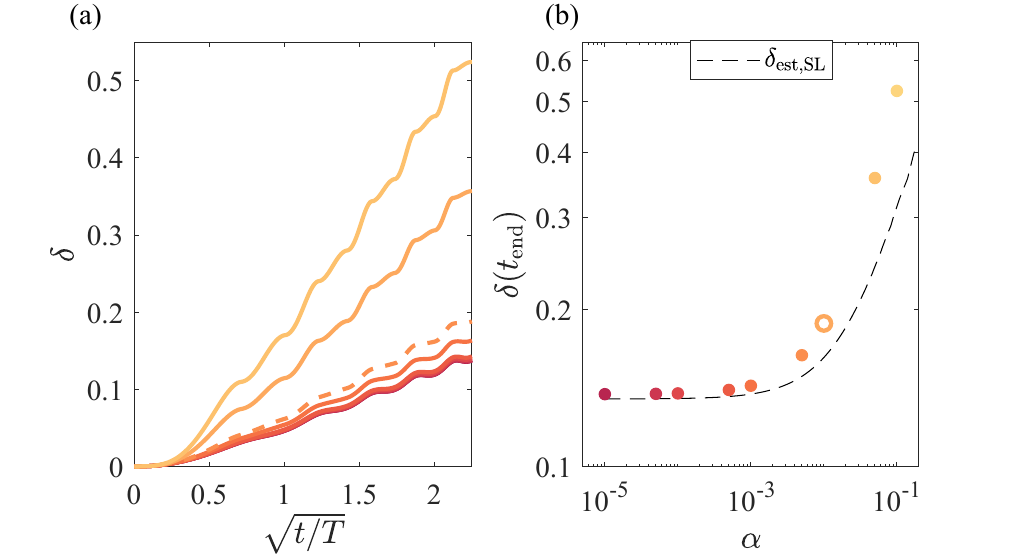}
    \caption{ Impact of $\alpha$ on the evolution of $\delta$ over 5 loading cycles. (a) We plot the evolution of $\delta$ with $\sqrt{t}$ for nine different values of $\alpha$ $\in [10^{-5},10^{-1}]$ (dark to light).  Note that the curves for the two smallest values of $\alpha$ overlap. In each case, delta is roughly linear in $\sqrt{t}$  with a slope that increases monotonically with $\alpha$. The dashed curve indicates the baseline value of $\alpha$. (b) We plot the final value of $\delta$ at $t=5T$ as function of $\alpha$. The open circle indicates the baseline value of $\alpha$. The black dashed curve is our estimate $\delta_\mathrm{est, SL}$ from equation~\eqref{delta-estimate-sl}. \label{fig:alpha} }
\end{figure}

We next isolate the roles of $\mathrm{Pe}^{-1}$ and $\alpha$. For that purpose, we focus on five cycles and consider a wide range of $\mathrm{Pe}^{-1}$ and $\alpha$. We consider the roles of $\phi_{f,0}$ and $l$ in Appendix~\ref{appendix-phi0-l}. The impact of changing $\mathrm{Pe}^{-1}$ and $\alpha$ on $\delta$ are shown in figures~\ref{fig:D0} and \ref{fig:alpha}, respectively.
All of the curves are roughly linear in $\sqrt{t}$ after an initial transient (\textit{i.e.}, spreading is Fickian on average), exhibiting fluctuations with a period $T$ because of the phenomena described in section~\ref{isolation-mechanisms}: loading decreases the porosity, forcing the solute to spread (advection to the left), and unloading reverses this process. Larger values of $\mathrm{Pe}^{-1}$ and $\alpha$ enhance the diffusive and dispersive fluxes, respectively, and hence drive faster spreading, as should be expected. For sufficiently small $\mathrm{Pe}^{-1}$, dispersion dominates diffusion and the rate of spreading becomes independent of $\mathrm{Pe}^{-1}$ (figure~\ref{fig:D0}). Similarly, diffusion dominates dispersion for sufficiently small $\alpha$ and the rate of spreading becomes independent of $\alpha$ (figure~\ref{fig:alpha}).

We next introduce a naive estimate for the travel distance during slow loading, $\delta_\mathrm{est, SL}$, based on the assumption that net transport is Fickian on average. That is, we assume that $\delta_\mathrm{est,SL} \propto\sqrt{\mathcal{D}_\mathrm{eff}t}$ for some effective diffusion/dispersion coefficient $\mathcal{D}_\mathrm{eff}\approx{}\mathrm{Pe}^{-1}+\alpha|v_f-v_s|$. During slow loading, the quantity $|v_f-v_s|$ is proportional to $A/T$ and decreases linearly from left to right~\citep[see][]{Fiori2023}. Thus, we assume that, on average, $|v_f-v_s|\sim{}C_2A/T$ for some constant $C_2$. The resulting estimate is then
\begin{equation}\label{delta-estimate-sl}
    \delta_\mathrm{est, SL} =C_1 f\left(\frac{\barphi}{\phi_{f,0}},\mathrm{Pe}^{-1}\right) \sqrt{ 4 \left(\mathrm{Pe}^{-1} + C_2 \alpha\frac{A}{T}\right) t},
\end{equation}
where $C_1$ is a constant and the function $f(\barphi/\phi_{f,0},\mathrm{Pe}^{-1})$ is an empirical prefactor to capture the impact of the average deformation on diffusive spreading, as discussed in more detail below (see figure~\ref{fig:phi_avg}), with $\barphi$ the overall average porosity (see equation~\ref{phi_avg}). By fitting equation~\eqref{delta-estimate-sl} to the results shown in figures~\ref{fig:D0}, \ref{fig:alpha}, and \ref{fig:phi_avg}, we find $C_1=1.29$, $C_2=0.25$, and 
\begin{equation}\label{function_phi_pe}
    % f\left(\frac{\barphi}{\phi_{f,0}} ,\mathrm{Pe}^{-1}\right) =\left( \frac{\barphi}{\phi_{f,0}} \right)^{0.2\mathrm{Pe}^{0.18}}.
    f\left(\frac{\barphi}{\phi_{f,0}},\mathrm{Pe}^{-1}\right)=1+C_3\ln(\mathrm{Pe}^{-1})\left(1-\frac{\barphi}{\phi_{f,0}}\right)
\end{equation}
with $C_3=0.123$. We discuss this functional form in detail below, around figure~\ref{fig:phi_avg}. With $\delta_\mathrm{est,SL}$ fully specified, we compare this prediction with the results in figures~\ref{fig:D0} and \ref{fig:alpha} (dashed black curves), where it provides a reasonable qualitative and quantitative estimate of $\delta$ across four orders of magnitude in both $\alpha$ and $\mathrm{Pe}^{-1}$. Note that the function $f$ is a constant of order 1 in figure~\ref{fig:alpha} because it does not depend on $\alpha$, whereas $f$ varies by a few percent across the full range of $\mathrm{Pe}^{-1}$ in figure~\ref{fig:D0} because it is a weak function of $\mathrm{Pe}^{-1}$.

This estimate ignores the periodic velocity field by assuming that dispersion occurs according to the time-averaged magnitude of $|v_f-v_s|$, our estimate for which is based on slow loading. During faster loading, $|v_f-v_s|$ will be increasingly localised near the left boundary and suppressed in the interior of the material. We consider the impact of localisation in the next section.

\subsection{Quantitative impacts of periodic loading on solute transport}\label{ST-different-regimes}

We next consider the effects of the loading parameters $A$ and $T$. To help interpret these results, we first consider the poromechanical response.

\subsubsection{Poromechanical response to periodic loading}\label{summary-paper1}

In our companion study~\citep{Fiori2023}, we explored the impact of $A$ and $T$ on the poromechanical response. During slow loading ($T\gg1$), the timescale of the loading is much slower than the poroelastic response of  the material and the response is quasi-static for any amplitude. The porosity is uniform in space throughout the cycle, returning to its undeformed value at the end of each cycle. The displacement, the fluid velocity, the solid velocity, and the Darcy flux all decrease linearly from a spatial extremum at the piston to zero at the right boundary. During fast loading ($T\ll1$), the timescale of the loading is much faster than the poroelastic response of the material. As a result, the deformation is non-uniform in space and increasingly localised near the left (permeable) boundary as the period decreases. The left portion of the domain experiences both compression and tension, whereas the right portion is in compression at all times. For very fast loading ($T\lll1$), the deformation is entirely localised near the left boundary and decays exponentially with $x$, such that the right portion of the material is in static compression. As the amplitude of the deformation increases, the change in porosity at the left boundary with respect to the relaxed state becomes increasingly asymmetric between loading and unloading, with a larger decrease (compression) during loading than the respective increase (tension) during unloading. 

\begin{figure}[tp]
    \centering
    \includegraphics[width=0.95\textwidth, trim=1cm 0.6cm 1cm 0cm]{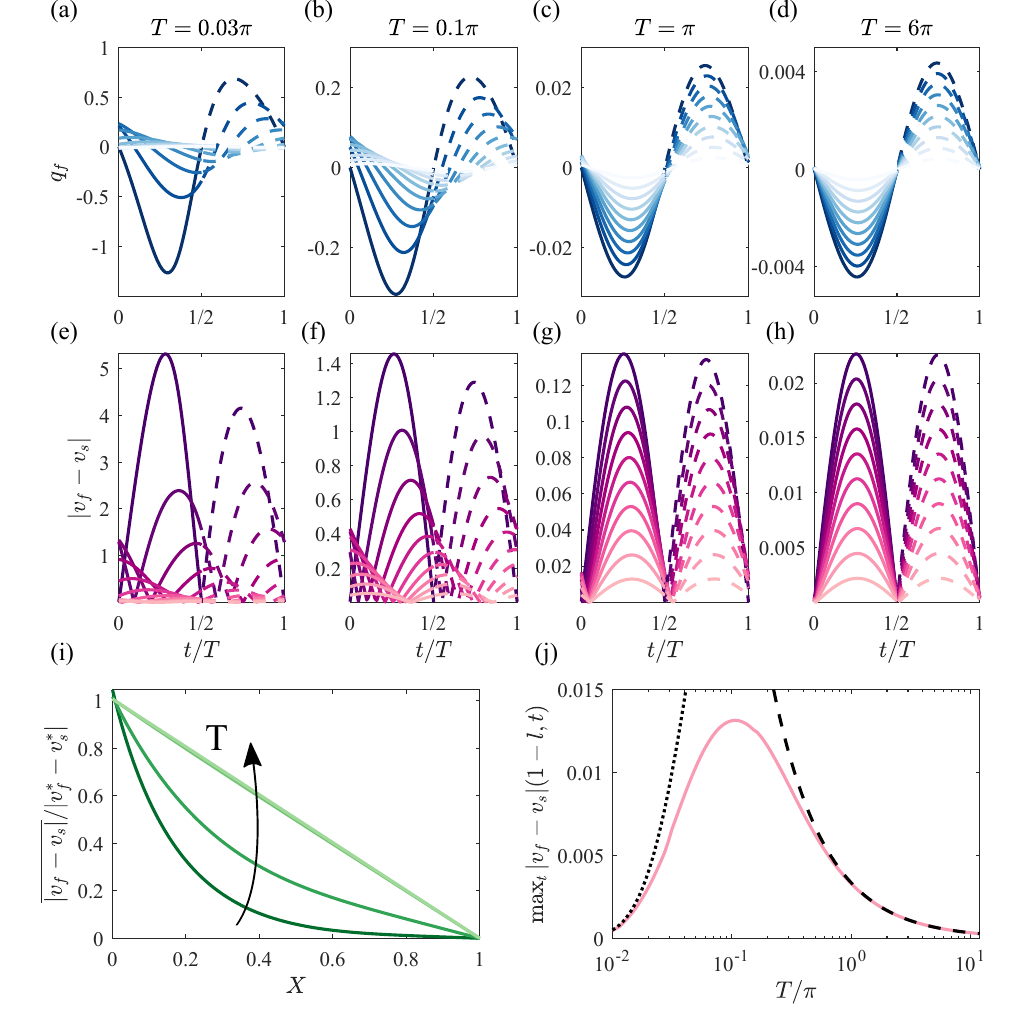}
    \caption{Evolution of (a)--(d)~fluid flux $q_f$ and (e)--(h)~$|v_f-v_s|$ at ten different values of $X=x-u_s(X,t)$ from $0$ to $1$ (dark to light) during one cycle for four different values of $T$ (columns). We distinguish between the loading half of the cycle ($\dot{a}>0$; solid curves) and the unloading half of the cycle ($\dot{a}<0$; dashed curves). (i)~Normalised time-average of $|v_f-v_s|$ as a function of $X$ for the same four values of $T$ (increasing dark to light). (j)~Maximum in time of $|v_f - v_s|$ at $x=1-l$ as a function of $T/\pi$ for $A=0.05$. The dashed black curve shows the slow-loading prediction of $\pi{}Al/(\phi_{f,0}T)$ and the dotted black curve shows the very-fast-loading prediction of $[\pi{}A/(\phi_{f,0}T)]\exp[-(1-l)\sqrt{\pi/T}]$ (see equation~\eqref{veryfast}). \label{fig:flux-T} }
\end{figure}

\begin{figure}[tp]
    \centering
    \includegraphics[width=0.95\textwidth, trim=1cm 0.6cm 1cm 0cm]{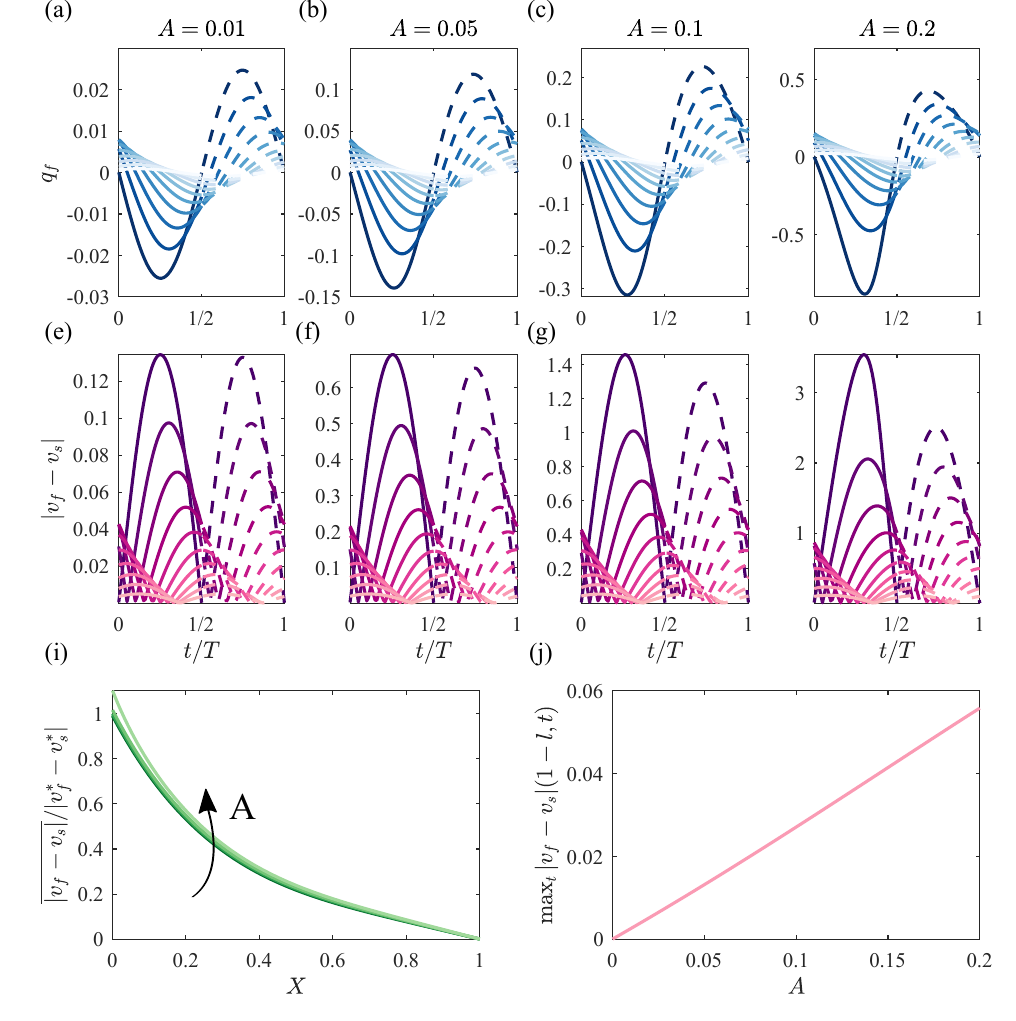}
    \caption{ Evolution of (a)--(d)~fluid flux $q_f$ and (e)--(h)~relative velocity $|v_f-v_s|$ at ten different values of $X=x-u_s(X,t)$ from $0$ to $1$ (dark to light) during one cycle for $T=0.1\pi$, and for four different values of $A$ (columns). We distinguish between the loading half of the cycle ($\dot{a}>0$; solid curves) and the unloading half of the cycle ($\dot{a}<0$; dashed curves). (i)~Normalised time-average of $|v_f-v_s|$ as function of $X$ for the same four values of $A$ (increasing dark to light). (j)~Maximum in time of $|v_f-v_s|$ at $x=1-l$ as a function of $A$. \label{fig:flux-A} }
\end{figure}

As noted above, the poromechanical response impacts solute transport through the motion of the fluid and through the changes in porosity. In figure~\ref{fig:flux-T} and figure~\ref{fig:flux-A}, respectively, we show the impact of $T$ and $A$ on the fluid flux $q_f$ (driving advection) and on the relative velocity $|v_f-v_s|$ (driving dispersion). As should be expected, the fluid flux exhibits localisation for fast loading and asymmetry in loading and unloading for large amplitudes.
In figure~\ref{fig:flux-T}, we fix $A$ to the baseline value and consider four values of $T$. For slow loading, $|v_f - v_s|\sim 2A(1-x)/T$ (see Ref.~\cite{Fiori2023}). For very large values of $T$ (\textit{e.g.}, figure~\ref{fig:flux-T}d and h), the deformation is uniform and very slow, and $|v_f - v_s|$ is low, especially toward the right (lighter shades) where the solute is positioned. Diffusion dominates over dispersion, even with a large amplitude. As $T$ decreases from $6\pi$ to $0.1\pi$ (\textit{e.g.}, figure~\ref{fig:flux-T}b,c,f,g), $|v_f - v_s|$ increases throughout the domain. As $T$ decreases further, however, the deformation is increasingly localised near the left boundary: figure~\ref{fig:flux-T}a and e show that both $q_f$ and $|v_f-v_s|$ are orders of magnitude larger near the left boundary (darkest curves) than near the right boundary (lightest curves). This localisation is highlighted in figure~\ref{fig:flux-T}i, where we plot the time-averaged profile of $|v_f - v_s|$ for each $T$, and in figure~\ref{fig:flux-T}j, where we plot the maximum value of $|v_f - v_s|$ at $x=1-l$ against $T/\pi$. Near the right boundary, where the solute is located, the relative velocity increases and then decreases with $T$, exhibiting a maximum around $T=0.1\pi$.

In figure~\ref{fig:flux-A}, we show the impact of $A$ on the same quantities for a fixed period, $T=0.1\pi$. The magnitudes of $q_f$ and $|v_f-v_s|$ increase monotonically with $A$ (figure~\ref{fig:flux-A}a--h) and the normalised time-averaged value of $|v_f-v_s|$ is relatively insensitive to $A$ (figure~\ref{fig:flux-A}i), suggesting that $|v_f-v_s|$ is essentially proportional $A$. This suggestion is confirmed in figure~\ref{fig:flux-A}j.

\subsubsection{Solute transport for different loading amplitudes and periods}

\begin{figure}[]    
    \centering
    \includegraphics[width=1\textwidth]{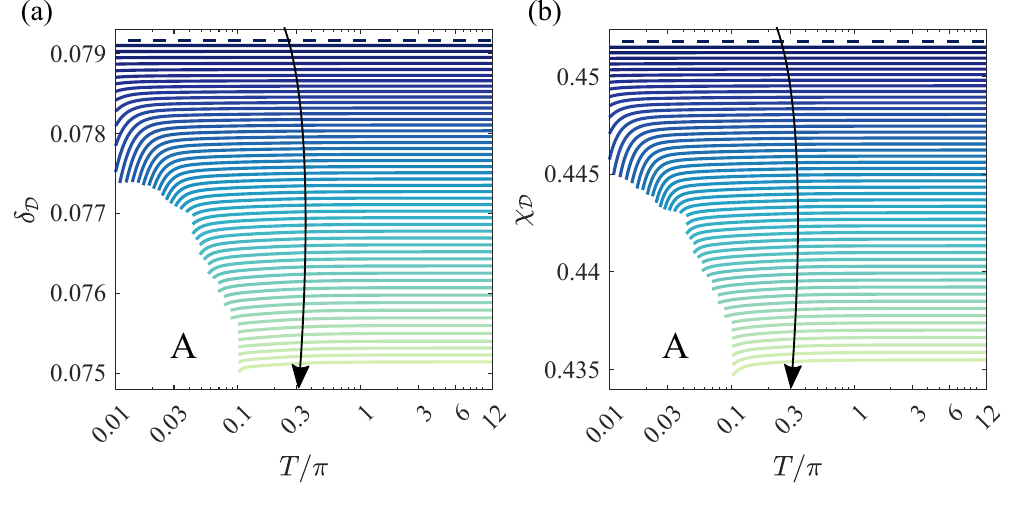}
    \caption{ (a)~Travel distance $\delta_\mathcal{D}$ and (b)~degree of mixing $\chi_\mathcal{D}$ as a function of $T$ and for a wide range of $A$ (evenly spaced from 0.004 to 0.2, increasing dark to light) after a loading time of $12\pi$ followed by a relaxation time of 1 (total time $12\pi+1$) for advection and molecular diffusion but no dispersion ($\alpha=0$). Note that the values of $T$ are selected to provide an integer number of loading cycles in a total time of $12\pi$, but the results are plotted as continuous curves for visual clarity. This constraint leads to periods ranging from $T=0.01\pi$ applied for 1200 cycles to $T=12\pi$ applied for 1 cycle. 
    Dashed lines (darkest color) correspond to diffusion with no loading ($A=0$). Note that the minimum porosity in the domain occurs near the piston and decreases monotonically with increasing $A$ and with decreasing $T$. Each curve ends on the left at the value of $T$ for which the minimum porosity vanishes and the simulations fail (see section~\ref{elasticity_skeleton}). \label{fig:delta-chi-diff} }
\end{figure}

We showed in section~\ref{isolation-mechanisms} how the three transport mechanisms act individually on the concentration profile. We now extend this analysis to examine the roles of $A$ and $T$. In figure~\ref{fig:delta-chi-diff}, we show the travel distance $\delta_\mathcal{D}$ and the degree of mixing $\chi_\mathcal{D}$ after a fixed total loading time of $12\pi$ followed by a relaxation time of 1 (total time $12\pi+1$), for advection and molecular diffusion but no dispersion ($\alpha=0$). We include results over a wide range of $T$ --- from very fast to slow loading --- and $A$ --- from small to large deformations.

We illustrated in section~\ref{isolation-mechanisms} that the contribution of advection is reversible over one loading cycle, independent of $A$ and $T$. However, as noted above, both the porosity field and the concentration gradients do depend on $A$ and $T$. Hence, molecular diffusion is expected to vary weakly with $A$ and $T$. The porosity $\phi_f$ is on average lower than the initial value $\phi_{f,0}$, because the loading has a non-zero mean --- the material is on average compressed. As noted in Ref.~\cite{Fiori2023}, the overall average porosity $\barphi$ over any integer number of cycles is given by
\begin{equation}\label{phi_avg}
    \barphi = \frac{1}{mT}\int_{nT}^{(n+m)T} \langle{\phi_f}\rangle \,\mathrm{d}t = 1 - \frac{1-\phi_{f,0}}{\sqrt{1-A}}, 
\end{equation}
where $m$ is an arbitrary positive integer and $n$ is an arbitrary non-negative integer. Note that $\barphi$ decreases with $A$ but is independent of $T$, as is also true of $\delta_\mathcal{D}$ and $\chi_\mathcal{D}$ in figure~\ref{fig:delta-chi-diff}, except for very fast loading. For slow loading, it is straightforward to show that the deformation stretches concentration gradients by, on average, a factor of $\phi_{f,0}/\barphi$.

\begin{figure}
    \centering
    \includegraphics{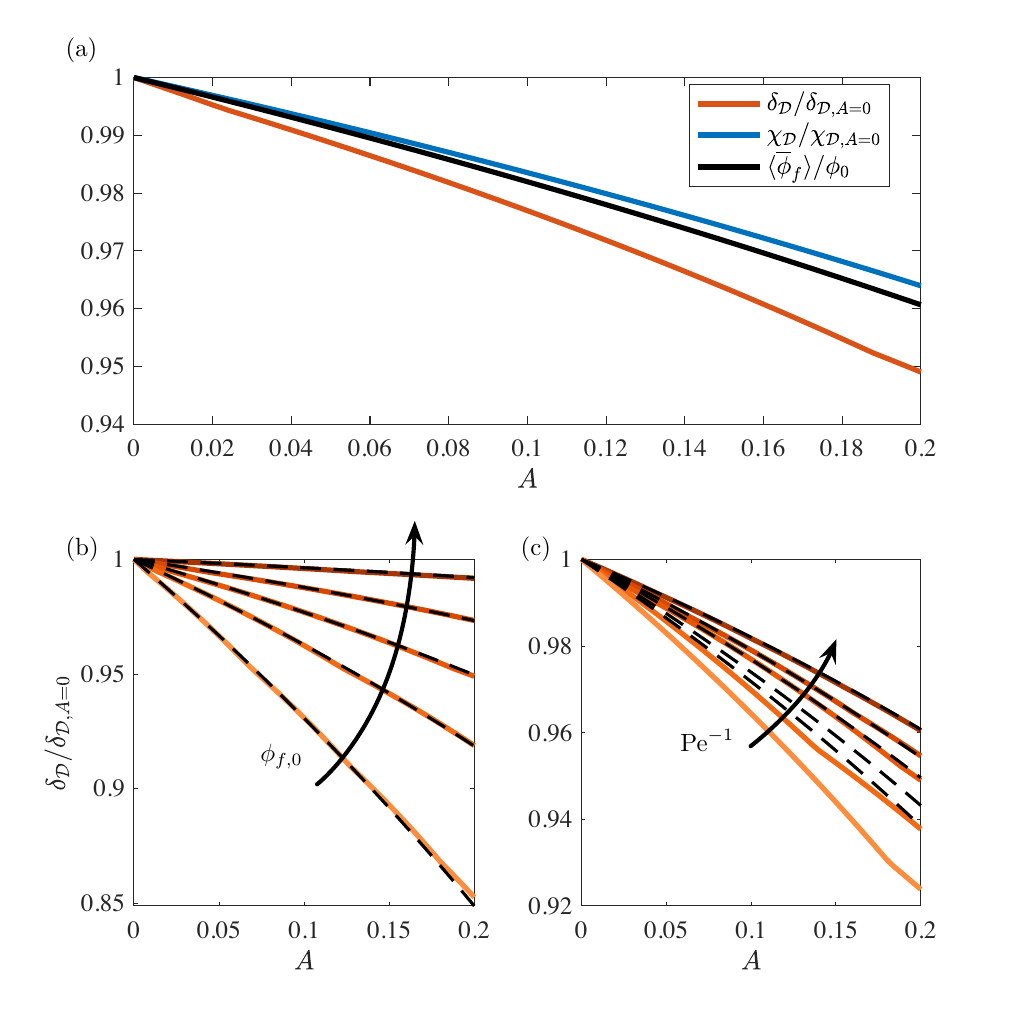}
    \caption{ (a)~Variation of $\delta_\mathcal{D}$ (red), $\chi_\mathcal{D}$ (blue), and $\barphi$ (black) with $A$ for $T=12\pi$, and where all three quantities are normalised by their values at $A=0$. We also plot the variation of $\delta_\mathcal{D}$ with $A$, again normalised by its value at $A=0$, for (b)~five values of $\phi_{f,0}$$\in[0.5,0.95]$ (light to dark) and (c)~five values of $\mathrm{Pe}^{-1}$$\in [3 \times 10^{-6},3 \times 10^{-4}]$ (light to dark). Black dashed lines are the empirical function $f(\barphi/\phi_{f,0},\mathrm{Pe}^{-1})$ from $\delta_\mathrm{est,SL}$ (see eqs.~\eqref{delta-estimate-sl} and \eqref{function_phi_pe}).  \label{fig:phi_avg} }
\end{figure}

In figure~\ref{fig:phi_avg}a, we fix $T=12\pi$ and plot $\delta_\mathcal{D}$, $\chi_\mathcal{D}$, and $\barphi$ against $A$; all three quantities are normalised by their values at $A=0$, demonstrating that they exhibit a qualitatively similar decay with $A$. We then plot $\delta_\mathcal{D}/\delta_{\mathcal{D}, A=0}$ against $A$ for several values of $\phi_{f,0}$ (figure~\ref{fig:phi_avg}b) and $\mathrm{Pe}^{-1}$ (figure~\ref{fig:phi_avg}c). Panel~(b) shows that increasing $\phi_{f,0}$ strongly mitigates the decay with $A$, while panel~(c) shows that increasing $\mathrm{Pe}^{-1}$ has a similar but much weaker effect.

With no dispersion ($\alpha=0$), the ratio of $\delta_\mathrm{est,SL}$ to its value for $A=0$ is precisely $f\left(\barphi/\phi_{f,0},\mathrm{Pe}^{-1}\right)$. Thus, the curves in figures~\ref{fig:phi_avg}b,c correspond to profiles of $f$ against $A$ for different values of $\phi_{f,0}$ and $\mathrm{Pe}^{-1}$. As indicated by the functional dependence of $f$, we hypothesise that $f$ depends on both $A$ and $\phi_{f,0}$ exclusively through the quantity $\barphi/\phi_{f,0}$, which, as noted above, measures the (inverse of the) average stretching of concentration gradients due to the deformation (see section~\ref{isolation-mechanisms}). As indicated in Equation~\eqref{function_phi_pe}, we find that $f$ is linear in $\barphi/\phi_{f,0}$ to a very close approximation; Figure~\ref{fig:phi_avg}b demonstrates excellent agreement between this expression for $f$ and our numerical results at fixed $\mathrm{Pe}^{-1}$.

The dependence of $f$ on $\mathrm{Pe}^{-1}$ is weaker. As indicated in Equation~\eqref{function_phi_pe} and illustrated in Figure~\ref{fig:phi_avg}c, we find that a logarithmic variation with $\mathrm{Pe}^{-1}$ provides reasonable qualitative agreement between our expression for $f$ and our numerical results, with good quantitative agreement for larger values of $\mathrm{Pe}^{-1}$. Note that our expression for $f$ has just one fitting parameter, $C_3$, the value of which is strongly constrained by the functional form of $f$ and by the clear linear trend with $\barphi/\phi_{f,0}$; of course, a better fit is possible with a different functional form and more fitting parameters. 

Although the precise functional form chosen here for $f$ is ultimately ad hoc, it is clear that stronger stretching of the concentration profile (\textit{i.e.}, increasing $A$ or $\phi_{f,0}$, decreasing $\barphi/\phi_{f,0}$) will more strongly hinder diffusion (decreasing $f$). The role of $\mathrm{Pe}^{-1}$ is more difficult to rationalise. We find that stretching of concentration gradients more strongly hinders diffusion at smaller values of $\mathrm{Pe}^{-1}$, corresponding to weaker diffusion. At smaller $\mathrm{Pe}^{-1}$, the initially steep concentration gradient will decay more slowly and it may be the case that diffusion is more sensitive to the stretching of these steeper gradients than shallower ones.

\begin{figure}
    \centering
    \includegraphics[width=0.9\textwidth]{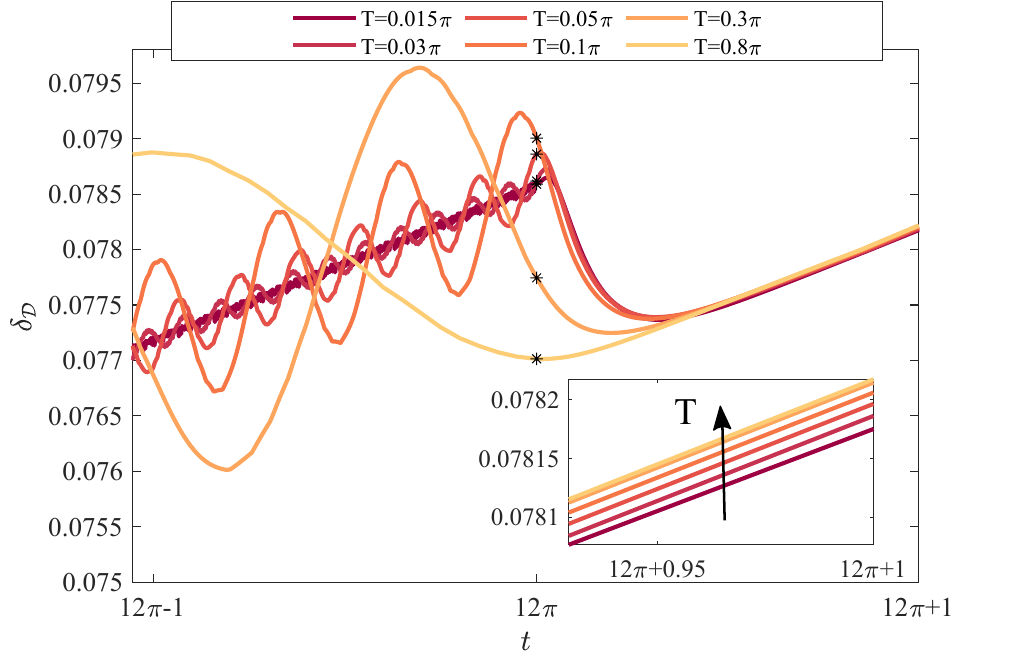}
    \caption{ Time evolution of $\delta_\mathcal{D}$ at the end of a long series of cycles for six of the smallest values of $T$ considered here, $T=0.015\pi$ to $0.8\pi$ (dark to light) with advection and molecular diffusion but no dispersion ($\alpha=0$). The inset focuses on the very last portion of the main plot. Black stars mark the end of the last cycle of periodic loading and the beginning of the relaxation phase (total time of 1), during which the material returns to its undeformed state. Note that the horizontal axis is on a log scale. \label{fig:deltas-t-diff} }
\end{figure}

Figures~\ref{fig:delta-chi-diff} and \ref{fig:phi_avg} confirm that $\delta_\mathcal{D}$ and $\chi_\mathcal{D}$ decrease weakly but monotonically with $A$, and are essentially independent of $T$ for all but the smallest values of $T$. For those smallest values, both $\delta_\mathcal{D}$ and $\chi_\mathcal{D}$ increase with $T$. This effect is not related to $\barphi$, which is independent of $T$ (see eq.~\eqref{phi_avg}). We explore this behaviour in figure~\ref{fig:deltas-t-diff} by plotting the evolution of $\delta_\mathcal{D}$ over time for a fixed amplitude $A=0.06$ and for several values of $T$, from $T=0.015\pi$ to $T=0.8\pi$.
For the slowest case ($T=0.8\pi$), $\delta_\mathcal{D}$ is minimum at the end of the loading time ($12\pi$, marked by a black star) and then increases due to diffusion during the relaxation time (a further time of 1). For smaller values of $T$, $\delta_\mathcal{D}$ instead decreases during the initial stages of relaxation, immediately after the end of loading. This decrease is due to the relaxation of the static far-field compression induced for very fast loading (\textit{e.g.},  $T \lesssim 0.1 \pi$). Since the deformation is much faster than the material response, the domain is never fully relaxed during periodic loading and the right portion, in particular, is in a state of static compression that contributes to an additional stretch of the concentration gradients that hinders diffusion. Once the loading stops, this compressed material relaxes by drawing fluid in, thereby further retracting the solute profile to a degree that is stronger and longer as $T$ decreases, such that $\delta_\mathcal{D}$ decreases as $T$ decreases in very fast loading (figures \ref{fig:delta-chi-diff} and \ref{fig:deltas-t-diff}).

\begin{figure}[tbp]
    \centering
    \includegraphics[width=1\textwidth]{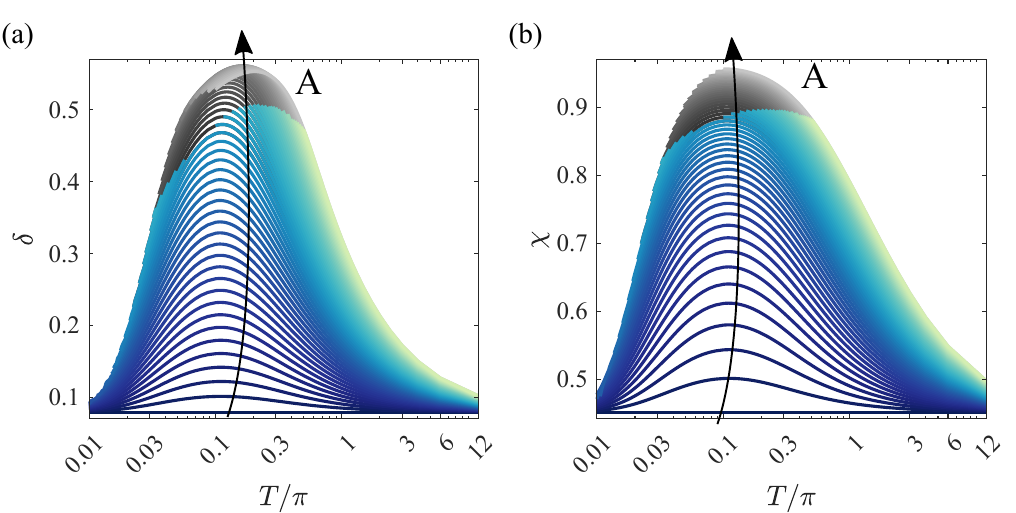}
    \caption{ (a)~Travel distance $\delta$ and (b)~degree of mixing $\chi$ as functions of $T$ and for a wide range of $A$ (increasing dark to light) after a loading time of $12\pi$ followed by a relaxation time of 1 (total time $12\pi+1$), as in figure~\ref{fig:delta-chi-diff}, but now  with all three transport mechanisms simultaneously active. The ranges of amplitudes and periods are the same as in figure~\ref{fig:delta-chi-diff}. Portions shown in greyscale indicate simulations where the solute reaches the left boundary and begins to leave the domain. \label{fig:delta-chi-maps} }
\end{figure}

\begin{figure}
    \centering
    \includegraphics[width=1\textwidth]{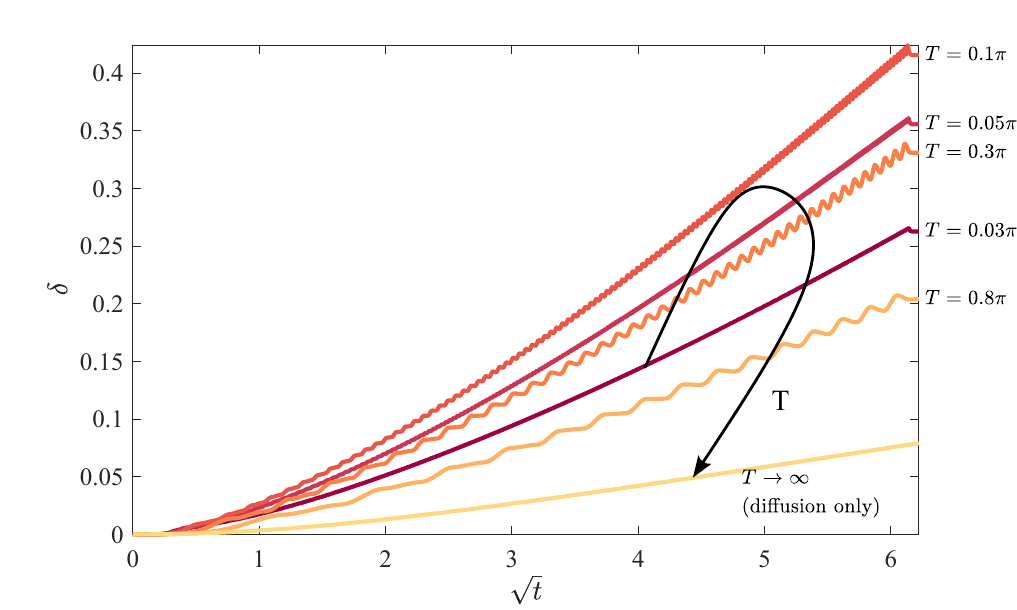}
    \caption{ Evolution of $\delta$ over the entire loading time for $A=0.06$ and for five values of $T$ (dark to light, values as indicated) when advection, molecular diffusion and hydrodynamic dispersion are simultaneously active. We also show the case of diffusion only (no loading, lightest curve). \label{fig:delta-local} }
\end{figure}

We next repeat the preceding analysis, but now including hydrodynamic dispersion (figure~\ref{fig:delta-chi-maps}). Both $\delta$ and $\chi$ are greatly enhanced by dispersion relative to the results in figure~\ref{fig:delta-chi-diff} across much of the range of $T$. Recall that the strength of dispersion is expected to scale with $A/T$. For very slow loading ($T\gtrsim{}12\pi$), the contribution of dispersion is negligible relative to that of diffusion and the values of $\delta$ and $\chi$ converge toward their values without dispersion. As $T$ decreases, the contribution of dispersion grows and increasingly dominates over diffusion, reaching a peak around $T\approx0.1\pi$. For these parameters, the values of $\delta$ in the peak region are almost one order of magnitude larger than without dispersion (cf. figure~\ref{fig:delta-chi-diff}). Across the full range of $T$ where dispersion dominates, $\delta$ and $\chi$ increase with $A$, as should be expected from figure~\ref{fig:flux-A}(i): The relative velocity in the interior increases monotonically with $A$. As $T$ further decreases, $\delta$ and $\chi$ instead begin to decrease and for very fast loading ($T\lesssim{}0.01\pi$) approach their values without dispersion. This trend can be linked to the impact of $T$ on the ratio of $|v_f - v_s|$ at $x=1-l$, shown in figure~\ref{fig:flux-T}(i): for faster loading, the deformation is increasingly localised near the piston and the region occupied by the solutes is increasingly not engaged. We examine these observations in more detail in figure~\ref{fig:delta-local}, by plotting the evolution of $\delta$ throughout the loading time for a fixed amplitude and for several periods between $T=0.03\pi$ and $T=0.8\pi$, thus spanning the peak in figure~\ref{fig:delta-chi-maps}. In all cases, $\delta$ exhibits oscillations with period $T$ on top of a roughly Fickian growth. As $T$ decreases, these oscillations decrease in magnitude as they increase in frequency, consistent with the deformation being increasingly localised at the left. The overall rate of spreading increases as $T$ decreases from $0.8\pi$ to $0.1\pi$, for which the increase in frequency leads to a net increase in dispersive flux despite the decrease in magnitude (see figure \ref{fig:flux-T}). As $T$ decreases further, the decrease in local magnitude begins to dominate the increase in frequency and the rate of spreading instead slows (see again figure \ref{fig:flux-T}).

We next compare the numerical values of $\delta$ shown in figure~\ref{fig:delta-chi-maps} with a revised estimate $\delta_\mathrm{est}$ that accounts for diffusion, dispersion, and localisation.
\begin{figure}   
    \centering
    \includegraphics[width=0.55\textwidth]{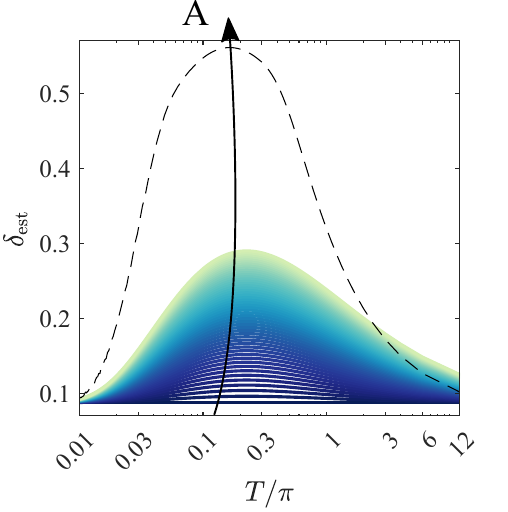}
    \caption{ Estimated travel distance $\delta_\mathrm{est}$ as a function of $T$ and for a wide range of $A$ (increasing dark to light) after a loading time of $12\pi$ followed by a relaxation time of 1 (total time $12\pi+1$), as in figure~\ref{fig:delta-chi-maps}, but here calculated via equation \eqref{delta_est_fl}. The dashed black line corresponds to the maxima for $\delta$ from figure~\ref{fig:delta-chi-maps}, for comparison. \label{fig:overA} }
\end{figure}
An analytical expression for the Darcy flux during very fast loading at very low amplitude ($T\ll1$ and $A\ll1$; from linear poroelasticity and analogous to Stokes's classical ``second problem'') can be derived from the results of Ref.~\cite{Fiori2023} by calculating the solid velocity as $v_{s,\mathrm{vf}}\approx{}\mathrm{d}u_{s,\mathrm{vf}}/\mathrm{d}t$, where the solid displacement $u_{s,\mathrm{vf}}$ is given in their equation~(3.12) and the subscript (``$\mathrm{vf}$'') stands for ``very fast''. Equation~\eqref{vf-vs} (also their equation~2.20) implies that the Darcy flux is then $\phi_{f,\mathrm{vf}}(v_{f,\mathrm{vf}}-v_{s,\mathrm{vf}})=-v_{s,\mathrm{vf}}$, such that
\begin{equation}\label{veryfast}
\begin{split}
    \phi_{f,\mathrm{vf}} (v_{f,\mathrm{vf}} -v_{s,\mathrm{vf}})  =
    - &\frac{\pi A}{T}\left[ \exp\left(-x\sqrt\frac{\pi}{T}\right)\sin\left(\frac{2\pi{}t}{T}-x\sqrt\frac{\pi}{T}\right)\right].
\end{split}
\end{equation} 
Equation~\eqref{veryfast} suggests that, in this regime, the amplitude of the Darcy flux decays exponentially with distance from the left boundary and with $T^{-1/2}$. Based on this solution, we reformulate equation \eqref{delta-estimate-sl} to include this exponential localisation in the magnitude of the dispersive flux:
\begin{equation}\label{delta_est_fl}
    \delta_\mathrm{est}=C_1 f\left(\frac{\barphi}{\phi_{f,0}} ,\mathrm{Pe}^{-1}\right)  \sqrt{  4 \left[\mathrm{Pe}^{-1} + C_2 \alpha \frac{A}{T}  \exp\left( {- (1-l) \sqrt{\frac{\pi}{T}}}\right)\right] t}, 
\end{equation}
where $f\left(\frac{\barphi}{\phi_{f,0}} ,\mathrm{Pe}^{-1}\right)$ is as defined in equation~\eqref{function_phi_pe}. Figure~\ref{fig:overA} shows that $\delta_\mathrm{est}$ captures the qualitative trends observed for $\delta$ in figure~\ref{fig:delta-chi-maps} (for which we report the maxima for a direct comparison), with the central peak that is however lower and slightly shifted compared to the one observed for $\delta$. We do not expect strong quantitative agreement because this estimate ignores the details of the spatial variation in the dispersive flux, which decreases linearly from left to right even for slow loading, as well as the oscillatory nature of the flow, and, most importantly, is based on an analytical solution that is only valid for small deformations and in the very-fast-loading region of the parameter space.

Hence, figures~\ref{fig:delta-chi-maps}--\ref{fig:overA} highlight two competing mechanisms: progressively faster loading enhances dispersion by promoting large dispersive fluxes in general, but also progressively localises the flow and deformation near the left boundary, suppressing the flux (and thus dispersion) in the interior. The competition between these two behaviours is the origin of the local maximum in $\delta$ and $\chi$ with $T$ and is qualitatively captured by the estimated travel distance $\delta_\mathrm{est}$.

In summary, this analysis reveals the link between transport fluxes and loading parameters and, consequently, the poromechanics of periodic loading. Through the travel distance $\delta_\mathrm{est}$, we are able to capture the three main transport regimes, corresponding to analogous regions of the $A-T$ domain shown in Figure~\ref{fig:delta-chi-maps}: 
\begin{itemize}
    \item $T\gtrsim 0.3\pi$: the loading is slow and the strength of dispersion is roughly proportional to $A/T$. Thus, faster and larger deformations progressively increase the strength of dispersion relative to diffusion. In particular, $|v_f-v_s|$ reaches higher peaks as $T$ decreases, promoting hydrodynamic dispersion, which is the dominant mechanism for this region.
    \item $0.1 \pi \lesssim T \lesssim 0.3 \pi$: the loading is fast, with an optimal balance between magnitude and depth of the fluid fluxes; dispersion is at its peak.
    \item $T \lesssim 0.1 \pi$: the loading is fast and the strong localisation of the flow and deformation near the left boundary dominates, increasingly suppressing dispersion in the interior as $T$ further decreases.
\end{itemize}
Note that the quantitative values of $\delta$ and $\chi$ also depend on the specific values of $\mathrm{Pe}^{-1}$ and $\alpha$, but the evolution of the porosity and fluid velocity with $A$ and $T$ does not. As a result, varying $\mathrm{Pe}^{-1}$ and $\alpha$ would change the width and height of the curves in figure~\ref{fig:delta-chi-maps}, but would not change the qualitative features of these plots or the position of the central peak.

\subsection{P{\'{e}}clet number in biological examples}

We conclude by reporting typical dimensionless loading parameters and diffusion coefficients for several examples of soft biological tissues (table~\ref{table-peclet}). Based on these values, we calculate the P{\'{e}}clet number and the effective P{\'{e}}clet number, as defined in section~\ref{scaling}, to understand the range of loading scenarios where advection and dispersion are likely to be non-negligible. For simplicity, we use a moderate dimensionless amplitude $\tilde{A}=0.1$ in all cases. The detailed parameters to calculate $\tilde{T}$ and $\mathrm{Pe}$ are reported, for the same biological tissues, in \cite{Fiori2023}. In all cases, the values of $\mathrm{Pe_{eff}}$ range from moderate to high, indicating that advection and potentially dispersion are unlikely to be negligible in these systems.

\begin{table}
    \caption{Loading and transport parameters for some examples of biological materials. \label{table-peclet} }
    \begin{center}
        \begin{tabular}{l|c|c|c|c|c}
            \hline
            Tissue & $\mathcal{D}_m$ [$\mathrm{m}^2\,\mathrm{s}^{-1}$] & $\mathrm{Pe}$ & $\tilde{T}/\pi$ & $\mathrm{Pe_{eff}}=\mathrm{Pe}(\tilde{A}/\tilde{T})$ \\
            \hline
            Brain ECM \citep{kedarasetti-fbcns-2022} & $1.4\times10^{-10}$ & $30$ & $0.003$ -- $0.1$ & $3\times10^{1}$ -- $1\times{}10^3$ \\
            \hline
            Cartilage \citep{Ferguson2004, gardiner2007, didomenico2017antibody}  & $5\times10^{-12}$ & $7\times10^3$ & \makecell{ $0.03$ -- $3$ (sitting) \\ $0.003$ -- $0.03$ (running) } & \makecell{ $10^2$ -- $10^3$ \\ $10^3$ -- $10^5$ } \\
            \hline
            \makecell{ Intervertebral disc \\ (Annulus F.) \citep{Ferguson2004} } & $5\times10^{-12}$ & $4\times10^2$ & \makecell{ $0.3$ (wake cycle) \\ $6\times10^{-5}$ -- $6\times10^{-3}$ (sitting) \\ $6\times10^{-6}$ -- $6\times10^{-5}$ (running) } & \makecell{ $4\times10^{1}$ -- $2\times10^3$ \\ $2\times10^3$ -- $2\times10^5$ \\ $2\times10^5$ -- $2\times10^6$ } \\
            \hline
            \makecell{ Cartilage scaffold \\ (bioreactor) \citep{Sengers2004} }  & $10^{-9}$ & $10^4$ & $10^{-4}$ -- $10^{-1}$ & $4\times10^{1}$ -- $4\times10^4$ \\
            \hline
        \end{tabular}
    \end{center}
\end{table}

\section{Conclusions}

We have derived physical insight into solute transport and mixing in a periodically compressed soft porous material under large deformations. To do so, we used a 1D continuum model, formulated following a rigorous nonlinear kinematic approach and considering Hencky elasticity as the constitutive law for the solid skeleton. Overall, we demonstrated that the characteristic rearranging of the porous structure --- resulting in a strong coupling between mechanical loading and fluid flow --- originates phenomena that are not reversible, despite the cyclic nature of the applied load.

Our analysis is linked to a companion study, where we characterise the mechanical response of a soft porous medium under the same loading scenarios \cite{Fiori2023}. That study showed that, depending on the loading period, the deformation can belong to either a slow-loading regime -- in which the deformation is uniform throughout the domain -- or a fast-loading regime -- in which the deformation is increasingly localised near the permeable boundary (\textit{i.e.}, the left boundary) \cite{Fiori2023}. Here, we analysed how these different mechanical behaviours related to the loading parameters influence solute transport and mixing. 

We first studied the nature of the transport phenomena, which act in the same direction during loading  --- from right to left --- whereas only advection changes direction during unloading (figure~\ref{fig:fluxes}). 
Next, we focused on a single loading cycle and compared the evolution of the solute concentration profile for four cases: (1) molecular diffusion only (no loading); (2) advection only (no diffusion or dispersion); (3) advection and diffusion only (no dispersion); and (4) a general case where all the transport mechanisms are active. We found that advection is reversible at the end of each loading cycle, and that deformation weakly suppresses diffusion by decreasing the porosity and by stretching the concentration gradients. 

We then considered two variables --- travel distance $\delta$ and degree of mixing $\chi$ --- and compared them for different transport coefficients and different loading parameters over several loading cycles. We showed that diffusion and dispersion are roughly Fickian on average. For advection and diffusion only (no dispersion), larger amplitudes increasingly reduce the average porosity and hence increasingly suppress diffusion. This case is essentially independent of $T$ except for very small values of $T$, where the material is additionally compressed and $\delta$ and $\chi$ decrease as $T$ decreases. When including dispersion, solute transport increases monotonically with $A$ but varies non-monotonically with $T$. The latter is due to the progressive localisation of the deformation at the left boundary, which reduces the intensity of the dispersive flux at the right boundary, where the solutes are positioned. We formalise these trends in the expression of the characteristic transport length $\delta_{est}$, which summarises the different regimes according to the transport and loading parameters.

For small deformations, it may be possible to derive an effective equation for net transport in this system. For example, \citet{Pool2016} used a multiple-scales technique to develop an effective transport equation for periodic loading in a semi-infinite domain that is valid for times much larger than the forcing period. The same approach could potentially be adapted to the present scenario, where the small-deformation analytical solution for the flow field is a full Fourier series rather than a single exponentially damped mode because the domain is bounded. This analysis is beyond the scope of the present work, but it would provide valuable qualitative insight into deformation-driven transport.

Our results have two important applications: the prediction of concentration profiles and the control of concentration profiles. The former is useful in cases where the conditions are fixed (\textit{e.g.}, to predict nutrient distributions in biological tissues \textit{in vivo}), whereas the latter could enable the optimization of the stimulation and material features to enhance or suppress spreading and mixing, or to reach a desired final solute profile (\textit{e.g.}, when designing scaffolds and bioreactors for tissue engineering).

We have taken the diffusion and dispersion coefficients themselves to be constants. As noted above, future work should consider the impact of the changing pore structure on the values of these material properties and, ultimately, on the underlying model for hydrodynamic dispersion. At the least, it is likely that, much like $k$, both $D_m$ and $\alpha$ should be deformation-dependent for moderate to large deformations.

Future work should also include the expansion of this study to 2D and 3D media with heterogeneous properties. Already for small deformations, periodic deformations have been shown to originate complex transport and mixing dynamics in heterogeneous media~\cite{pool2018, Trefry2019, Wu2020, Wu2024} and in response to multi-modal forcing~\citep{Trefry2020}. Complex media and forcings are particularly relevant to biological systems; examples include heterogeneous structures originating in the multi-scale nature of tissues and/or multi-modal loading originating in the co-existence of different vital cyclic pulsations, such as the cardiac and the respiratory cycles. Finally, future work should also focus on an experimental validation of the results presented here.

Declaration of interests: The authors report no conflicts of interest.

\begin{acknowledgments}
This work was supported by the European Research Council (ERC) under the European Union's Horizon 2020 Programme [Grant No. 805469]. S.P. was supported by Start-Up Research Grant (SRG/2021/001269) by the Science and Engineering Research Board, Department of Science and Technology, Government of India. For the purpose of Open Access, the authors have applied a CC BY public copyright licence to any Author Accepted Manuscript (AAM) version arising from this submission. We thank Luis Cueto-Felgueroso for the helpful discussion on compact finite differences. 
\end{acknowledgments}

%\clearpage 

\appendix 
\section{Numerical method} \label{numerical-method}

% We solve our model using compact finite-differences in space and implicit Runge-Kutta integration in time. We achieve the latter using \texttt{MATLAB}'s built-in solver \texttt{ODE15s}~\cite{shampine-siamjscicomput-1997}.

Our model equations are solved using finite difference methods. 
We use a sixth-order compact finite difference approximation for spatial derivatives and a third-order compact finite difference approximation for derivative boundary conditions~\cite{lele1992compact}. For time integration, we use an implicit Runge-Kutta method via \texttt{MATLAB}'s built-in solver \texttt{ODE15s}~\cite{shampine-siamjscicomput-1997}.

To account for the moving boundary, we work in a scaled coordinate system. We start from a general conservation law of the form: 
\begin{equation}\label{pde-xt}
\frac{\partial \Phi(x,t)}{\partial t}+ \frac{\partial}{\partial x}[F(\Phi(x,t)]=0.
\end{equation}
We then rescale the spatial coordinate $x$, introducing the variable
\begin{equation}
    \xi=\frac{x-a}{1-a},
\end{equation}
which maps $a\leq{}x\leq{}1$ to $0\leq{}\xi\leq{}1$. Performing a standard change of variables from $(x,t)$ to $(\xi,t)$, partial derivatives transform according to
\begin{equation}
    \frac{\partial}{\partial{t}} \,\to\, \frac{\partial}{\partial{t}}+\frac{\partial{\xi}}{\partial{t}}\frac{\partial}{\partial \xi} \quad\mathrm{and}\quad \frac{\partial}{\partial{x}} \,\to\, \frac{\partial{\xi}}{\partial x}\frac{\partial}{\partial \xi}
\end{equation}
and equation~\eqref{pde-xt} can be rewritten as
\begin{equation}
    \frac{\partial \Phi}{\partial t}-
\frac{(1-\xi) }{(1-a)}\dot{a} \frac{\partial \Phi}{\partial \xi} +\bigg(\frac{1}{1-a}\bigg) \frac{\partial F( \Phi)}{\partial \xi}=0.
\end{equation}
When solving equation~\eqref{conservation-q-scaled}, we then take $\Phi=\phi_f$ and
\begin{equation}
    {F(\phi_f)}=-\tilde{D}_f(\phi_f)\frac{\partial{\phi_f}}{\partial{\tilde{x}}}, 
\end{equation}
while for equation~\ref{conservation-c-scaled} we take ${\Phi=\phi_f \tilde{c}}$ and 
\begin{equation}
    F(\phi_f \tilde{c})= (\phi_f \tilde{c}) \tilde{v}_f - \phi_f \tilde{\mathcal{D}}  \frac{\partial \tilde{c}}{\partial \tilde{x}}.
\end{equation}

\begin{figure}
    \centering
    \includegraphics[width=0.9\textwidth]{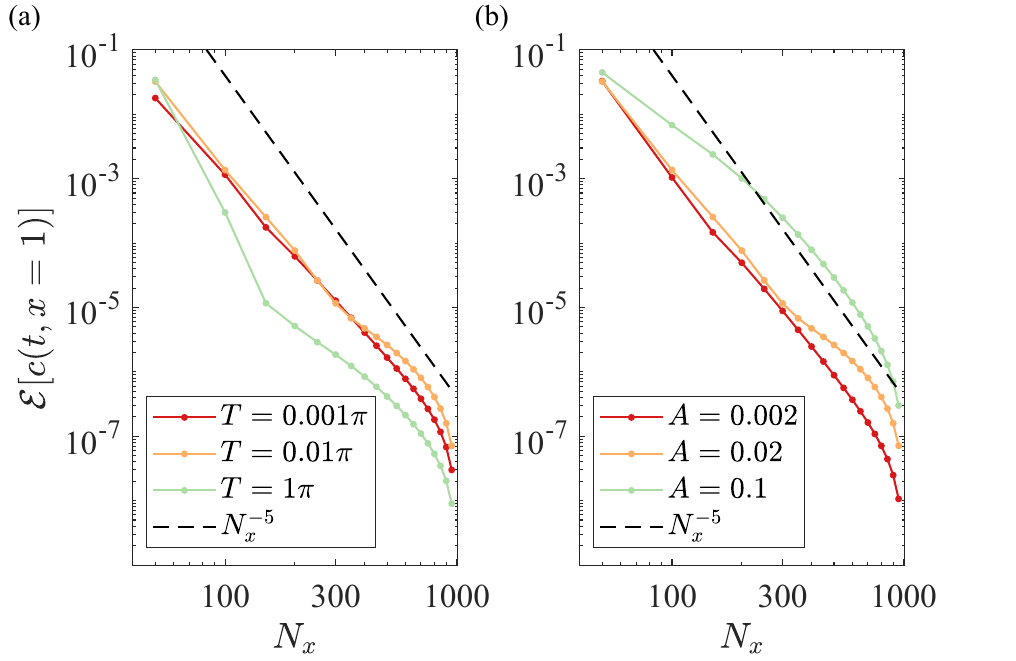}
    \caption{ Convergence analysis: RMS relative error in $c(t,x=1)$ relative to the solution for $N_x=1000$. (a) We fix $A=0.02$ and consider different values of $T$, from very fast to slow. (b) We fix $T=0.1\pi$ and consider different values of $A$, from small to large. \label{fig:nx} }
\end{figure}

For our spatial discretisation, we perform a convergence analysis in the number of grid points $N_x$ (see figure~\ref{fig:nx}) by calculating the Root Mean Square (RMS) relative error in $c(t,x=1)$ for each solution with respect to the solution for $N_x=1000$.
To balance between accuracy and computational time, we choose for all our simulations $N_x=300$, with a RMS error of less than $10^{-3}$.
We fix our absolute and relative error tolerances for time integration to be $10^{-10}$. 
 
As a reference, the cases of pure molecular diffusion and pure advection are compared with analytical solutions for this problem (see section~\ref{anl_dff}), resulting in good agreement between numerical and analytical results. 

\section{Analytical solution for molecular diffusion and advection}\label{anl_dff}

The analytical solution for the molecular diffusion of a step function in a semi-infinite material, as formulated by \citet{crank1979mathematics}, is:
\begin{equation}\label{diff_anl}
    c(x,t)=c_0 \bigg\{1- \frac{1}{2} \left[\mathrm{erfc}\left(\frac{l + x -1}{2\sqrt{\mathrm{Pe}^{-1} t}}\right) +\mathrm{erfc}\left(\frac{l - x +1}{2\sqrt{\mathrm{Pe}^{-1} t}}\right)\right]\bigg\}.
\end{equation}

\begin{figure}
    \centering
    \includegraphics[width=0.9\textwidth]{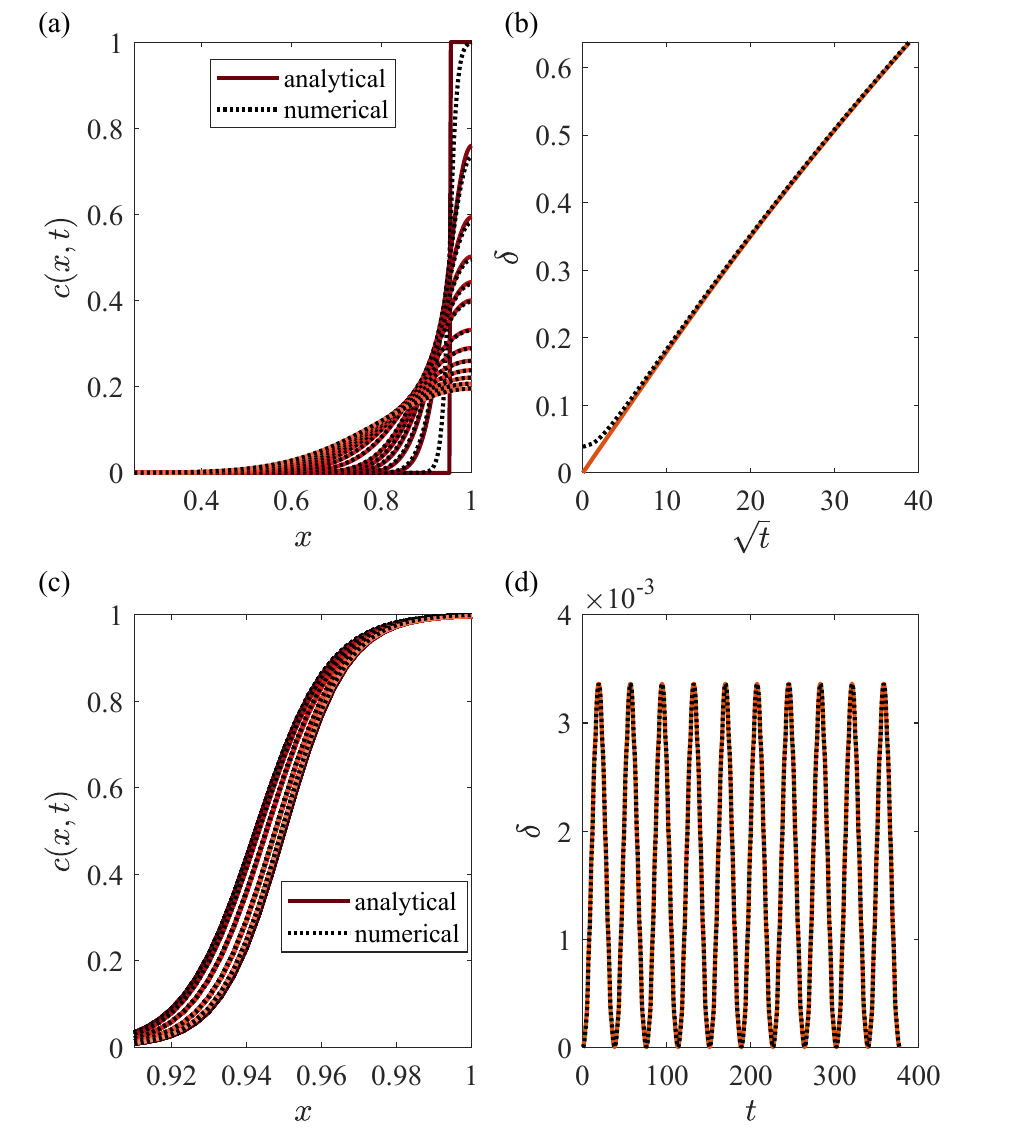}
    \caption{Qualitative comparison between the analytical solution (solid red curves) and numerical solution (dotted black curves) for diffusion only (a-b) and for advection (c-d). We show (a - c) the evolution of the concentration profiles $c(x,t)$ in time (dark to light) and (b-d) the evolution of $\delta$ over time.}
    \label{fig:analytical}
\end{figure}

For slow loading, the analytical solution for advection only ($\mathrm{Pe}^{-1}=\alpha=0$) is
\begin{equation}
    c({x},t)=\frac{1}{2} \bigg \{ \tanh{ \bigg[{s}\left( ({x}-1)\frac{\phi_{f,0}}{\phi_f}+{l}\right)}  \bigg]+1 \bigg \} .    
\end{equation}
In figure~\ref{fig:analytical}, we compare these analytical solutions to the numerical solutions obtained with the method described in Appendix~\ref{numerical-method}. Note that, for the case of diffusion, we consider an initial solute profile that is smoother than the one for which the analytical solution is formulated. Hence, the two solutions show an initial discrepancy that vanishes with time, as the profile adjusts towards classical self similarity.

\section{Dispersive flux}\label{appendix-dispersive}

\begin{figure}
    \centering
    \includegraphics[width=0.8\textwidth]{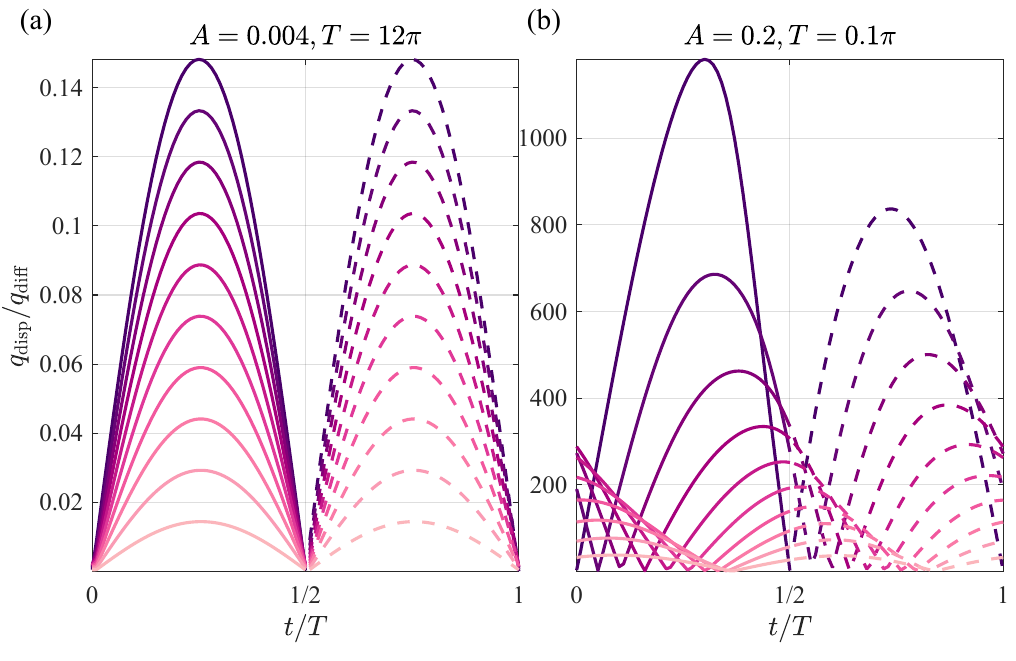}
    \caption{ (a)~Smallest and (b)~largest values of $q_\mathrm{disp}/q_\mathrm{diff}$ considered in this study. The flux is plot at ten different values of $X=x-u_s(X,t)$ from $0$ to $1$ (dark to light) during one cycle. We distinguish between the loading half of the cycle ($\dot{a}>0$; solid curves) and the unloading half of the cycle ($\dot{a}<0$; dashed curves). \label{fig:disp-range} }
\end{figure}

In this Appendix, we justify the choice of fixing $\mathrm{Pe}^{-1}$ and $\alpha$ to the specific baseline values reported in table~\ref{tab:quantitative}. We do so by quantifying the strength of dispersion relative to diffusion using the ratio of the dispersive solute flux $q_\mathrm{disp}$ to the diffusive solute flux $q_\mathrm{diff}$ which we define as 
\begin{equation}\label{qdisp_diff}
    \frac{q_\mathrm{disp}}{q_\mathrm{diff}}= \alpha \mathrm{Pe}| {v}_f-{v}_s|,
\end{equation}
and which measures the relative importance of these two mechanisms. 
In figure~\ref{fig:disp-range}, we compare  $q_\mathrm{disp}/q_\mathrm{diff}$ for the slow-loading and low-amplitude case and for the fast-loading and high-amplitude case. In the former, diffusion prevails over dispersion throughout the domain, while in the latter dispersion is the dominant mechanism. Hence, the specific baseline values chosen for $\mathrm{Pe}^{-1}$ and $\alpha$ are such that our results span the range from diffusion-dominated to dispersion-dominated transport across the range of $A$ and $T$ considered.  

\section{Initial porosity and solute strip width}\label{appendix-phi0-l}

\begin{figure}
    \centering
    \includegraphics[width=0.85\textwidth]{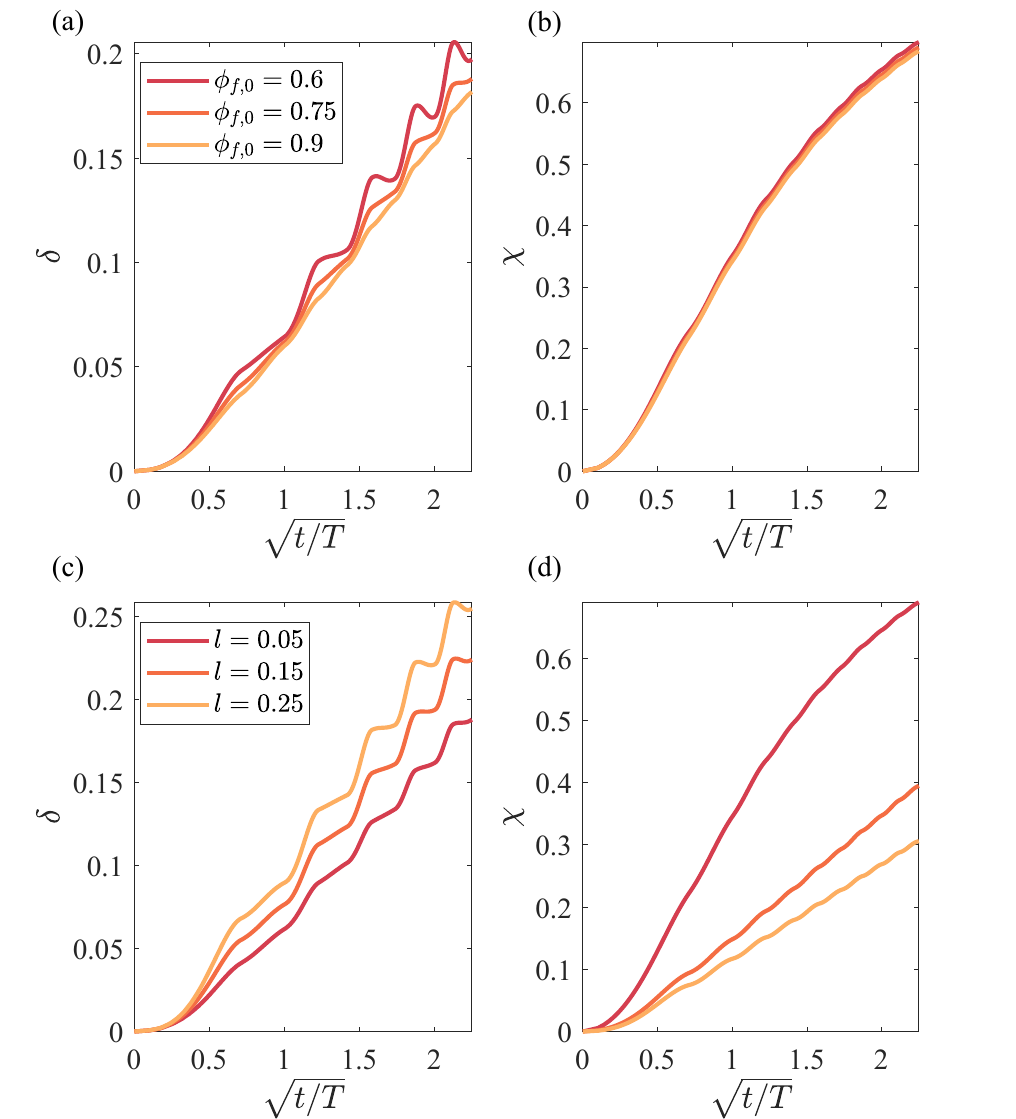}
    \caption{ Evolution of $\delta$ and $\chi$ over 5 cycles for different values for (a,b) the initial porosity $\phi_{f,0}$ and (c,d) the initial solute stripe width $l$. \label{fig:phi-l} }
\end{figure}

Results of varying $\phi_{f,0}$ and $l$ are shown in figure~\ref{fig:phi-l}. 
Decreasing $\phi_{f,0}$ leads to larger oscillations in fluid flux for a given solid velocity (see~\eqref{vf-vs}) In this way, both advection and hydrodynamic dispersion are enhanced. However, the resulting impact on travel distance in figure \ref{fig:phi-l} is relatively small because dispersion is relatively weak for these parameters.

Increasing the initial solute amount leads to a sharp increase in the travel distance: this is generally expected because the total concentration of solute is higher and therefore all the transport mechanisms are amplified. In addition, the fluid flux increases monotonically in magnitude from right to left, so larger values of $l$ expose the solute front to stronger advection and dispersion. However, note that a wider solute strip is associated with a lower degree of mixing because the variance of the solute compared to the initial variance is lower when there is more solute (see equation~\ref{eq:chi}). 

\clearpage

% \bibliographystyle{plainnat}
% \bibliography{references}

\end{document}